\documentclass[%
  aps,
  prl,
  reprint,
  superscriptaddress,
  amsmath,
  amssymb,
  longbibliography
]{revtex4-2}

\usepackage{graphicx}
\usepackage{bm}
\usepackage{xcolor}
\usepackage{booktabs}
\usepackage[hidelinks]{hyperref}

\begin{document}


\title{\textbf {TeV Higgsino Dark Matter from LZ Nuclear Recoil to Fermi-LAT Gamma Rays} 
}%

\author{Lei Wu}
\email{leiwu@njnu.edu.cn}
\affiliation{Department of Physics and Institute of Theoretical Physics, Nanjing Normal University, Nanjing, 210023, China}

\author{Yang Zhang}
\email{zhangyang2025@htu.edu.cn}
\affiliation{Center for Theoretical Physics, Henan Normal University, Xinxiang 453007, China}

\author{Bin Zhu}
\email{zhubin@mail.nankai.edu.cn}
\affiliation{School of Physics, Yantai University, Yantai 264005, China}

\date{\today}

\begin{abstract}
The excess of high-energy nuclear recoils, which was recently reported by the LUX-ZEPLIN (LZ) experiment, together with the Fermi-LAT Galactic-Center preference, suggests new physics beyond the standard model (SM). One of the minimal extensions of the SM is to introduce a fermionic electroweak doublet in which the lightest neutral component can play the role of a stable dark matter particle. We show that the minimal higgsino dark matter can offer a common origin for both excesses. The inferred mass splitting, however, depends on the local dark matter velocity distribution, and including the high-velocity tail generated by the gravitational influence of the Large Magellanic Cloud yields $\delta_m =481$--$519~\mathrm{keV}$ for $m_\chi =1000$--$1183~\mathrm{GeV}$ at $1\sigma$ level. Near elastic-scattering blind spots, solar-captured higgsinos fail to reach thermal equilibrium, weakening their annihilation signal, so that ten years of IceCube data exclude $\delta_m \lesssim 475~\mathrm{keV}$ at 90\% C.L. The viable parameter space can be tested through the associated gamma-ray line, which is within the projected reach of CTAO.

\end{abstract}

\maketitle

{\it Introduction.} The existence of dark matter (DM) is firmly established by gravitational evidence spanning galactic rotation curves, cluster lensing, the cosmic microwave background, and the growth of large-scale structure \cite{Bertone:2004pz}.  Its identity, however, remains one of the central open questions of particle physics and cosmology.  Weakly interacting massive particles (WIMPs) are among the most studied candidates, since masses and interactions at the weak scale, from a few tens of GeV to several TeV, reproduce the observed relic abundance and provide concrete targets for direct, indirect, and collider searches \cite{Jungman:1995df}. Recently, a single high-energy nuclear recoil in LUX-ZEPLIN and a gamma-ray excess toward the Galactic Center in Fermi-LAT data may point to the upper end of this mass range.

The LZ candidate is a single nuclear recoil at $E_R = 248 \pm 23_{\mathrm{stat}} \pm 23_{\mathrm{sys}}~\mathrm{keV_{nr}}$ and has a local significance of about $3.4\sigma$ and a global significance of $2.6\sigma$~\cite{LZ:2026axp}. This event is kinematically compatible with TeV-scale DM, though standard elastic scenarios are disfavored by null results in the low energy nuclear recoils. Meanwhile, the Fermi-LAT excess is a broad spectral component at multi-GeV energies, reported in 14 years of data with a significance of about $2\sigma$ \cite{Dessert:2022evk}. Its morphology follows a steeply rising, approximately squared Navarro--Frenk--White profile, and its spectrum is broader and harder than the diffuse emission that accounts for the rest of the Galactic plane, so that an interpretation in terms of unresolved astrophysical sources becomes increasingly strained at the highest energies.  A TeV-scale WIMP annihilating into gauge-boson and Higgs-boson pairs produces just such a component, since gauge-boson fragmentation yields a continuum peaking near $E_\gamma\sim m_\chi/50\simeq 20$~GeV and spanning about a decade in energy, matching both the position and the width of the excess. Both observations, despite their modest significance, are thus compatible with a common origin in a TeV-scale WIMP.

In this {\it Letter}, we propose minimal higgsino dark matter as an explanation of the above two intriguing excesses. In the pure-doublet limit, the observed thermal relic abundance fixes its mass near 1 TeV \cite{Arkani-Hamed:2006wnf,Jungman:1995df}, the very scale the two observations favor. It can inelastically scatter off xenon nuclei, reproducing the LZ recoil, and annihilate into $W^+W^-$ and $ZZ$, enhanced by the Sommerfeld effect at halo velocities \cite{Hisano:2004ds,Hisano:2006nn}, generating the continuum underlying the Fermi-LAT preference. Note that kinematics of endothermic inelastic scattering ties the 248 keV recoil to the mass splitting $\delta_m$ through the incident velocity, which must grow with $\delta_m$. The accessible $\delta_m$ is therefore bounded by the largest velocities available in the local dark matter distribution. We find that the high-velocity tail of this distribution, generated by the gravitational influence of the Large Magellanic Cloud (LMC)~\cite{Smith-Orlik:2023kyl}, yields the favored parameter space: $\delta_m =481$--$519$~keV for $m_\chi =1000$--$1183$~GeV at $1\sigma$ level. On the other hand, such a TeV-scale higgsino is captured in the Sun and subsequently annihilates into neutrinos. The large inelastic splitting that explains the LZ event makes inelastic scattering the dominant process governing the captured population, which therefore does not reach thermal equilibrium within the solar age. The resulting nonthermal distribution suppresses the annihilation rate relative to the equilibrium assumption and weakens the corresponding IceCube bounds. Simulating the nonthermal evolution, we find that ten years of IceCube data exclude $\delta_m \lesssim 475$~keV at 90\% C.L.. In addition, higgsino annihilation yields gamma-ray signals within the projected reach of Cherenkov Telescope Array Observatory (CTAO)~\cite{Rodd:2024qsi}.

{\it Methodology.} We consider a minimal higgsino dark matter that saturates the relic density. We denote the lightest-state mass by \(m_{\chi_1^0}\) and the neutral-state splitting by \(\delta_m\equiv m_{\chi_2^0}-m_{\chi_1^0}\). This small splitting enables the inelastic transition $\chi_1^0 A \to \chi_2^0 A$, which can produce a sizable nuclear recoil signal via the unsuppressed off-diagonal $Z$ coupling~\cite{Smirnov:2026aqk}. In the heavy gaugino limit, the leading contribution to the mass splitting is given by,
\begin{equation}
\delta_m \simeq
m_Z^2
\left|
s_W^2
\frac{M_1+\mu\sin 2\beta}{M_1^2-\mu^2}
+
c_W^2
\frac{M_2+\mu\sin 2\beta}{M_2^2-\mu^2}
\right|,
\label{eq:higgsino_splitting}
\end{equation}
The parameter $\mu$ mainly determines the higgsino mass and
thermal annihilation rate, while \(M_1\), \(M_2\), and
\(\tan\beta\) control the neutral splitting and residual elastic
interaction (see details in the Supplemental Material). A splitting of a few hundred keV can be realized in
two limiting ways. At finite gaugino masses, opposite-sign bino
and wino contributions can cancel, approximately along
\(M_1\simeq-\tan^2\theta_W M_2\). Alternatively, both
contributions can become small through common gaugino decoupling,
which for the splitting of interest requires gaugino masses of
order \(10^4~\mathrm{TeV}\).

For inelastic scattering on a nucleus \(A\), the minimum incident speed required to produce a recoil energy \(E_R\) is
\begin{equation}
v_{\min}(E_R)
=
\frac{m_AE_R/\mu_{\chi A}+\delta_m}
{\sqrt{2m_AE_R}},
\label{eq:vmin_inelastic}
\end{equation}
where \(\mu_{\chi A}\) is the dark-matter--nucleus reduced mass. A positive mass splitting raises the kinematic threshold and shifts the recoil spectrum toward high energies. For a thermal higgsino near \(1.1~\mathrm{TeV}\), xenon recoils in the
\(\mathcal{O}(100)~\mathrm{keV}\)
range probe neutralino mass splittings of several hundred keV. The corresponding event rate is controlled by the high-velocity tail of the Galactic dark matter distribution~\cite{Tucker-Smith:2001myb,Bramante:2016rdh,Graham:2024syw}. Once the relic abundance fixes \(m_\chi\), the LZ recoil spectrum therefore primarily selects the neutral-state splitting \(\delta_m\).

The same thermal higgsino also provides a predictive interpretation of the Fermi-LAT data. Its present-day annihilation rate and continuum photon spectrum are determined primarily by electroweak gauge interactions and \(m_\chi\), and are therefore much less sensitive to the gaugino parameters that control \(\delta_m\). Once the relic abundance restricts \(m_\chi\) to the TeV scale, the particle-physics input to the comparison with Fermi-LAT data is essentially fixed. The resulting thermal-higgsino template is compatible with the mild preference reported in the public fourteen-year Fermi-LAT Galactic-center analysis~\cite{Dessert:2022evk}. Fermi-LAT therefore provides an independent consistency test of the LZ interpretation rather than an additional determination of the neutral-state splitting.

\begin{table}[tbp]
\caption{\textbf{Joint-likelihood inputs and complementary tests.}}
\label{tab:likelihood-contract}
\footnotesize
\begin{ruledtabular}
\begin{tabular}{@{}p{0.14\columnwidth}p{0.40\columnwidth}p{0.34\columnwidth}@{}}
{\raggedright Term\par} &
{\raggedright Likelihood statistic/probe\par} &
{\raggedright Input\par}\\
\hline
{\raggedright Relic\par} &
{\raggedright $\displaystyle
\chi_\Omega^2=\left(\frac{\Omega_\chi h^2-\Omega_0}{\sigma_\Omega}\right)^2$
\par} &
{\raggedright $\Omega_0=0.120$;
$\sigma_\Omega=0.01206$~\cite{Planck:2018vyg}\par}\\[1.0ex]

{\raggedright LZ\par} &
{\raggedright \(q_{{\rm LZ}}\): one-event extended recoil-energy likelihood\par} &
{\raggedright \({\cal E}_{\rm LZ}=2.84~\mathrm{tonne\,yr}\) and
\(E_{\rm obs}=248~\mathrm{keV}\), with the effective background calibrated
to the public \(\mathcal O_1^s\) local significance. Profiled Maxwellian or
fixed Halo~13 MW+LMC mean template~\cite{LZ:2026axp,Smith-Orlik:2023kyl}\par}\\[1.0ex]

{\raggedright Fermi-LAT\par} &
{\raggedright $\displaystyle q_F=2\ln
\frac{{\cal L}^{\rm Ein}_F(m_\chi,\langle\sigma v\rangle_H)}
{{\cal L}^{\rm Ein}_F(m_\chi,0)}$\par} &
{\raggedright Non-Gaussian; the full released nine-annulus likelihood cube is
used~\cite{Dessert:2022evk}.\par}\\[1.0ex]
\hline

{\raggedright IceCube\par} &
{\raggedright Solar-neutrino search\par} &
{\raggedright Ten-year solar-WIMP limit~\cite{IceCube:2025fcu}\par}\\[0.5ex]

{\raggedright H.E.S.S.\par} &
{\raggedright Galactic-center line search\par} &
{\raggedright $546~\mathrm{h}$ observed line limits; two published
solar-distance choices~\cite{HESS:2026ila}\par}\\[0.5ex]

{\raggedright CTAO\par} &
{\raggedright Galactic-center $\gamma+X$ forecast\par} &
{\raggedright $500~\mathrm{h}$ Alpha line sensitivity; quoted Omega
gain~\cite{CTAO:2024wvb}\par}\\[0.5ex]

{\raggedright Dwarfs\par} &
{\raggedright Dwarf-galaxy $W^+W^-$ search\par} &
{\raggedright Five-instrument limits; two $J$-factor
treatments~\cite{Fermi-LAT:2025gei}\par}\\[0.5ex]

{\raggedright Planck\par} &
{\raggedright CMB energy injection\par} &
{\raggedright 2018 $p_{\rm ann}$ bound~\cite{Slatyer:2015jla,
Planck:2018vyg}}\par\\[0.5ex]

{\raggedright ATLAS\par} &
{\raggedright Disappearing-track search\par} &
{\raggedright $13~\mathrm{TeV}$ lifetime-dependent higgsino
limit~\cite{ATLAS:2026hnb}\par}\\
\end{tabular}
\end{ruledtabular}
\end{table}

We combine the relic-density, LZ, and Fermi-LAT likelihoods in Table~\ref{tab:likelihood-contract} to test whether a single thermal higgsino can accommodate both observations as follows,
\begin{equation}
\begin{aligned}
\chi_{{\rm joint}}^2
&\equiv 
\chi_\Omega^2(m_\chi)
+
q_{{\rm LZ}}(m_\chi,\delta_m)
-
q_F(m_\chi).
\end{aligned}
\label{eq:joint_likelihood}
\end{equation}
In our fitting, we consider halo treatments, a profiled Maxwellian model and the fixed mean velocity distribution of the Halo~13 MW+LMC analogue. The numerical inputs and detailed construction of the three contributions are given in the Supplemental Material. 

\begin{figure}[t]
\centering
\includegraphics[width=\columnwidth]{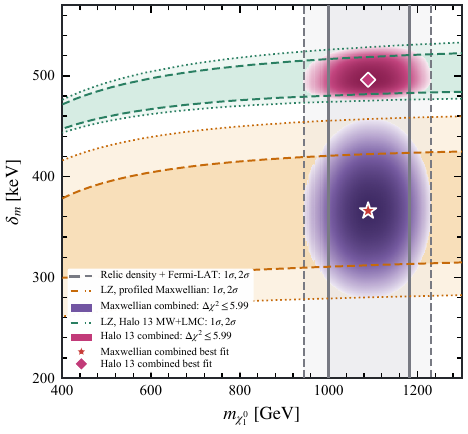}
\caption{ The $1\sigma$ and $2\sigma$ likelihood regions obtained from relic-density, Fermi-LAT, and LZ likelihood. For the relic-density plus Fermi-LAT contribution, the gray band with
solid and dashed boundaries shows \(\Delta\chi^2\leq2.30,5.99 \). The orange and green bands
show the corresponding LZ likelihood regions for the profiled Maxwellian and
Halo~13 MW+LMC treatments, respectively. The violet and magenta shadings show
the corresponding combined \(\Delta\chi_{{\rm joint}}^2\leq5.99\) regions, with the star and diamond marking the best-fit points.}
\label{fig:joint_likelihood}
\end{figure}

In Fig.~\ref{fig:joint_likelihood}, we show the likelihood-favored region on the plane of the higgsino dark matter mass $m_\chi$ and the mass splitting $\delta_m$ for two dark matter velocity distributions. As mentioned above, the relic density dominates the mass constraint and the LZ recoil selects the splitting. The Fermi-LAT contribution depends on the assumed Galactic-center density profile. For the FIRE-2 Romulus profile, the best fit among the simulated Milky Way analogues, the preferred annihilation cross section agrees with the thermal-higgsino prediction within $1\sigma$~\cite{Dessert:2022evk}, while for the fiducial Einasto profile adopted here it varies only weakly over the relic-selected mass range.

The relic-density
contribution fixes the mass projection, which gives \(1000\lesssim m_\chi/\mathrm{GeV}\lesssim1183\) at
\(1\sigma\) and
\(943\lesssim m_\chi/\mathrm{GeV}\lesssim1230\) at \(2\sigma\). The halo
choice instead acts mainly on the splitting. The profiled Maxwellian
likelihood peaks at \(m_\chi=1.09~\mathrm{TeV}\) and
\(\delta_m=366~\mathrm{keV}\), with
\(312\lesssim\delta_m/\mathrm{keV}
\lesssim422\) at \(1\sigma\) and
\(280\lesssim\delta_m/\mathrm{keV}
\lesssim457\) at \(2\sigma\).
The upper reach in
\(\delta_m\) is controlled primarily by the kinematic endpoint
\(v_{\rm esc}+v_E\), rather than by the signal normalization required to
reproduce the LZ candidate.
The LMC-enhanced high-velocity tail extends this
kinematic reach and therefore shifts the preferred splitting upward  to
\(\delta_m=481\)--\(519~\mathrm{keV}\)
and
\(\delta_m=475\)--\(529~\mathrm{keV}\),
respectively.

{\it Constraints and Prospects.}
Due to the same higgsino–nucleus interactions underlying the LZ recoil, higgsinos can be captured by the Sun, where their annihilation produces neutrinos~\cite{Press:1985ug,Spergel:1984re,Gould:1987ir,Widmark:2017yvd,Nussinov:2009ft,Menon:2009qj, Catena:2018vzc,Pospelov:2026ewn}. The scenario is therefore directly testable by IceCube searches for solar neutrinos~\cite{IceCube:2025fcu}.
The captured higgsinos initially lose orbital energy through inelastic scattering, while the subleading elastic interactions complete their thermalization.
However, the elastic cross sections depend sensitively on the small gaugino admixture of the higgsino-like LSP and receive important loop corrections \cite{Bisal:2024ezn}. The spin-independent cross section can become very small near the top--stop threshold, where the top--stop correction to the Higgs vertex allows the tree-level and vertex amplitudes to cancel the net twist-2 box and two-loop gluon contribution~\cite{Bisal:2026hpm}. Meanwhile, as shown in the
Supplemental Material, the renormalized \(Z\)-exchange and electroweak-box amplitudes have opposite signs, so varying the neutralino composition can strongly suppress the neutron spin-dependent cross section. Similar cancellations appear in proton spin-dependent cross section~\cite{Hisano:2004pv}.
Without sufficiently strong elastic scattering,
inelastic cooling eventually becomes kinematically forbidden before
thermalization is complete, leaving a diffuse population on extended solar
orbits~\cite{Blennow:2018xwu}. In this case, the neutrino
signal cannot be evaluated by assuming either a thermalized dark-matter
distribution or capture--annihilation equilibrium, and its strength is
substantially reduced relative to the thermalized-equilibrium estimate.

To evaluate this non-equilibrium signal, we simulate the
evolution of the captured higgsino population on solar orbits. The simulation
follows continuous capture, inelastic
transitions between orbit states, evaporation, and annihilation
simultaneously. Dividing the bound phase space into orbit
states \(g\), we evolve the number of orbits as,
\begin{align}
 \frac{\mathrm{d}N_g}{\mathrm{d}t}
 &=C_g+\sum_h\left(\Sigma_{hg}N_h-\Sigma_{gh}N_g\right)\nonumber\\
 &\quad-\mathcal{E}_gN_g-N_g\sum_h C_{A,gh}^{\rm tot}N_h .
 \label{eq:nonthermal-population}
\end{align}
Here \(C_g\) continuously injects captured particles into
orbit state \(g\). The two terms involving \(\Sigma_{gh}\) describe
inelastic transitions into and out of this state, \(\mathcal{E}_gN_g\) accounts for
evaporation, and the final term removes particles through annihilation with
all populated orbit states \(h\), with
\(C_{A,gh}^{\rm tot}=\sum_X C_{A,gh}^{X}\).
The orbit-pair coefficients \(C_{A,gh}^{X}\)
combine the spatial overlap of states \(g\) and \(h\) with the
relative-velocity-averaged annihilation kernel. The extended nonthermal
orbits reduce this overlap relative to a compact thermal distribution.
Unlike the capture--annihilation equilibrium limit, the signal also retains
an explicit dependence on the velocity-averaged annihilation cross section.
We therefore include the channel-dependent Sommerfeld enhancement
~\cite{Hisano:2004ds,Hisano:2006nn}, which gives
\(\langle\sigma v\rangle_{\rm eff}\simeq
1.5\times10^{-26}~\mathrm{cm^3\,s^{-1}}\) over the relevant orbit
velocities. The orbit overlaps and velocity convolution are detailed in the
Supplemental Material.
With these coefficients, we compute the
annihilation rate in each relevant final state \(X\) from the evolved orbit populations,
\begin{align}
 \Gamma_X=\frac{1}{2}\sum_{g,h}C_{A,gh}^{X}N_gN_h ,
 \label{eq:nonthermal-annihilation-rate}
\end{align}
and then map the \(WW+ZZ\) signal to the IceCube response ratio as
\begin{align}
 R_{WW+ZZ}^{\rm resp}=\frac{\Gamma_{WW}+\kappa_{ZZ}\Gamma_{ZZ}}
 {\Gamma_{A,WW}^{90}} .
 \label{eq:icecube-response-ratio}
\end{align}
For the IceCube response, we retain \(X=WW,ZZ\), and
\(\Gamma_{A,WW}^{90}=1.72\times10^{19}~\mathrm{s}^{-1}\)
is the ten-year IceCube pure-\(WW\) rate limit~\cite{IceCube:2025fcu}, while \(\kappa_{ZZ}=1.16\)
accounts for the relative \(ZZ\) response. Points with
\(R_{WW+ZZ}^{\rm resp} > 1\) are excluded by the IceCube data at 90\% C.L. The construction
of \(\kappa_{ZZ}\) is given in the Supplemental Material.

In Fig.~\ref{fig:icecube-exclusion}, we show the ten-year IceCube lower bound on the mass splitting $\delta_m$. When the elastic scattering vanishes, inelastic capture remains efficient, but the extended
nonthermal population has a smaller annihilation rate. The corresponding boundary falls to
\(\delta_m\gtrsim475~\mathrm{keV}\) at 90\% C.L. In contrast, if the elastic scattering is permitted, the capture--annihilation equilibrium may be reached, which produces
\(\delta_m\gtrsim553~\mathrm{keV}\) and thus lies outside the \(2\sigma\) regions favored by LZ. We also consider the uncertainties from the solar models. To maximize the reduction of the IceCube limit, we adopt the individual \(2\sigma\) lower abundances of Fe, Zn, Ge, Pb, and U within
the B23/SF-III AGSS09 solar structure
~\cite{herrera_2023_10174172,Asplund:2009fu,Asplund:2021solar}. Because capture at large \(\delta_m\) becomes
increasingly sensitive to heavy nuclei, especially Fe, the lower abundances move the IceCube boundary to
\(\delta_m\gtrsim466~\mathrm{keV}\), as shown in the loose solar/nuclear inputs scenario in Fig.~\ref{fig:icecube-exclusion}. The exact abundance rescalings and radial profiles are given in the Supplemental
Material. According to the shell-model calculations, we also note that there is an additional $\sim 10-20\%$ uncertainty from the nuclear-response of Fe~\cite{Catena:2015uha}, which could lower this boundary by about \(12~\mathrm{keV}\), to
\(454~\mathrm{keV}\).

For the profiled Maxwellian distribution, the LZ \(2\sigma\) upper edge at
\(\delta_m=457~\mathrm{keV}\) lies only about
\(9~\mathrm{keV}\) below the loose IceCube boundary at \(466~\mathrm{keV}\).
This small separation is
kinematic: the LZ upper edge is set by the endpoint of the fitted speed
distribution, rather than by an inability to reproduce the candidate event.
A stronger high-velocity tail can therefore shift the LZ-preferred splitting
above the IceCube boundary. Indeed, for the Halo~13 MW+LMC distribution, the
entire LZ \(2\sigma\) interval,
\(\delta_m=475\text{--}529~\mathrm{keV}\), lies
above the loose IceCube boundary at \(466~\mathrm{keV}\). At the Halo~13 MW+LMC
best-fit splitting, \(\delta_m\simeq496~\mathrm{keV}\), the loose prediction is
\(R_{WW+ZZ}^{\rm resp}\simeq0.065\), more than one order of magnitude below the IceCube limit.
Thus the full Halo~13 MW+LMC LZ \(2\sigma\) region can satisfy the IceCube bound for the blind spot of the elastic scattering.


\begin{figure}[t]
\centering
\includegraphics[width=\columnwidth]{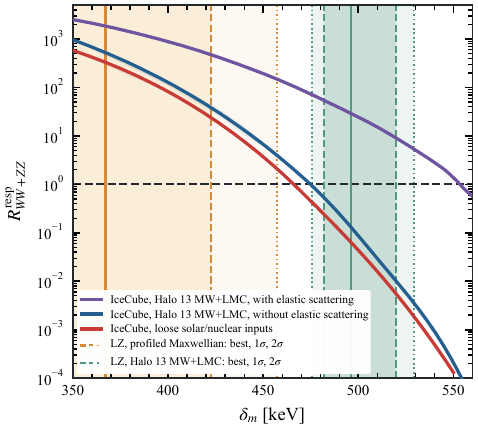}
\caption{IceCube constraints on solar-captured higgsino dark matter at $m_{\chi_1^0}=1.11~\mathrm{TeV}$. The blue curve shows the signal from the nonthermal population induced by the pure inelastic scattering. As a comparison, the results for the elastic scattering (purple curve) and loose solar/nuclear inputs (red curve) scenarios are displayed as well. The orange and green
bands show the LZ likelihood regions for the profiled Maxwellian and Halo~13 MW+LMC halo, respectively. The solid, dashed, and dotted vertical lines mark the best-fit splitting and the \(1\sigma\) and \(2\sigma\) LZ boundaries, respectively. }
\label{fig:icecube-exclusion}
\end{figure}

\begin{figure}[t]
\centering
\includegraphics[width=\columnwidth]{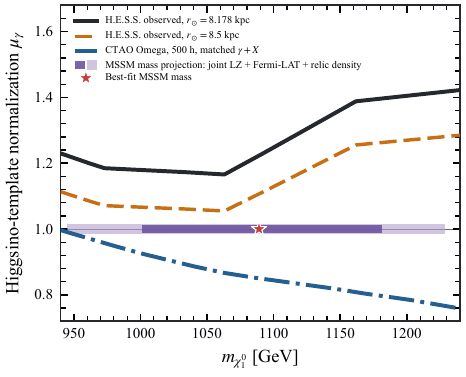}
\caption{Sensitivity of gamma rays from Galactic-center to thermal higgsino dark
matter. Here $\mu_\gamma=1$ is the full-density thermal prediction. The black
and orange curves are the H.E.S.S.~observed line-search limits expressed
relative to the canonical NLO higgsino signal for the two published choices of
\(R_0\), and the blue curve is the response-matched $500~\mathrm{h}$
CTAO Omega estimate for the complete higgsino $\gamma+X$ spectrum. The dark and
light violet bands are the $\Delta\chi_{\rm joint}^2\leq2.30$ and $5.99$ mass
projections, and the star marks the joint best-fit mass.}
\label{fig:hess-ctao}
\end{figure}

The parameter region selected by the joint analysis also
predicts a line-like Galactic-center signal from present-day higgsino
annihilation. Figure~\ref{fig:hess-ctao} compares
the current H.E.S.S. limits and the projected CTAO sensitivity with the mass
range selected by the joint analysis. A TeV higgsino produces
Sommerfeld-enhanced $\gamma\gamma$ and
$\gamma Z$ lines accompanied by a sizeable endpoint contribution~\cite{Beneke:2019gtg,
Urban:2021cdu,Beneke:2022eci}. H.E.S.S. searches for monochromatic gamma-ray
lines from the Galactic center and reports no significant excess~\cite{HESS:2026ila}.
We map its observed limits onto a higgsino signal-strength axis by
dividing them by the NLO Sommerfeld-enhanced higgsino line prediction
computed with \textsc{DM$\gamma$Spec}, so that
$\mu_\gamma=1$ denotes the full-density thermal prediction.  At the joint
best-fit mass, the thermal higgsino line prediction reaches $0.815$--$0.900$ of
the two observed limits and is therefore close to, but not excluded by, the
present H.E.S.S. limits.
Excluding the complete target region would require a
substantially stronger H.E.S.S. line limit. Across the
full $\Delta\chi_{\rm joint}^2\leq5.99$ mass projection, the ideal
background-dominated scaling $\mu_{\gamma,95}\propto T^{-1/2}$ implies total
exposures of about $900$ and $1100~\mathrm{h}$ for the two published choices
of the solar Galactocentric distance \(R_0\), respectively. The conservative
requirement is thus roughly
twice the current $546~\mathrm{h}$ exposure.

CTAO offers a more promising route to the thermal target because
its larger effective area and improved energy resolution at TeV energies
increase the sensitivity to sharp spectral features, while a higgsino-template
search can exploit the endpoint photons in addition to the two lines~\cite{CTAO:2024wvb}.
We compute the complete NLO higgsino $\gamma+X$ spectrum with
\textsc{DM$\gamma$Spec}~\cite{Beneke:2019gtg,Urban:2021cdu,Beneke:2022eci} and
derive the $500~\mathrm{h}$ Alpha projection by folding it through the public
instrument response and matching it to the released pure-line sensitivity.
The resulting projection gives $\mu_{\gamma,95}=1.69$ at
$m_\chi\simeq1.1~\mathrm{TeV}$ and remains above the thermal target.
However, the public
response corresponds to the initial Alpha array, and the CTAO study estimates that
the larger, full-scope Omega configuration improves the Alpha line limits by
about a factor of two~\cite{CTAO:2024wvb}. Since our higgsino projection is anchored to
the same pure-line sensitivity, we transfer this Alpha-to-Omega gain to the
full template, giving
$\mu_{\gamma,95}^{\rm Omega,est}=\mu_{\gamma,95}^{\rm Alpha,match}/2=0.85$.
This value lies below $\mu_\gamma=1$, placing the mass interval selected by the
joint analysis within exclusion reach if no signal is observed.
The H.E.S.S. recast and CTAO response matching are
detailed in the Supplemental Material.

Besides, current dwarf-galaxy~\cite{Fermi-LAT:2025gei}, CMB~\cite{Planck:2018vyg}, and disappearing-track searches at the LHC~\cite{ATLAS:2026hnb} do not exclude the selected thermal-higgsino region, with quantitative comparisons given in the Supplemental Material. Heavy-element paleodetectors provide a complementary probe of this inelastic-higgsino region and can reach mass splittings up to about \(920~\mathrm{keV}\)~\cite{Graham:2026ivn}. Future short-track searches at a 100-TeV hadron collider or a multi-TeV muon collider could directly probe the thermal higgsino~\cite{Mahbubani:2017gjh,Capdevilla:2024bwt}.

{\it Conclusions.} We have investigated a thermal higgsino interpretation of
the LZ high-energy recoil and the Fermi-LAT Galactic-center preference.
Thermal freeze-out fixes the higgsino mass near \(1.1~\mathrm{TeV}\), while
the LZ recoil selects the neutral-state splitting through the high-velocity
tail of the local dark-matter distribution. With the LMC-enhanced Halo~13
distribution, the joint \(1\sigma\) region gives
\(481\lesssim\delta_m/\mathrm{keV}\lesssim519\). Solar-neutrino searches
provide a decisive complementary test. A thermalizing captured population
requires \(\delta_m\gtrsim553~\mathrm{keV}\), outside the LZ \(2\sigma\)
regions. Near the blind spot of elastic scattering, however, the captured higgsinos remain
on extended nonthermal orbits, lowering the IceCube boundary to
\(475~\mathrm{keV}\), or to \(466~\mathrm{keV}\) with the loose solar and
nuclear inputs. The full Halo~13 LZ \(2\sigma\) interval can therefore remain
compatible with IceCube when the residual elastic scattering is negligible.
The same region predicts a Galactic-center line and endpoint signal close to
the present H.E.S.S. limits and within the projected reach of CTAO, providing
a direct test of this interpretation.

{\it Note added.}
Ref.~\cite{Rodd:2026tyn}
pointed out that the higgsino interpretation of the LZ event can predict
additional nuclear recoils in the high-energy sideband. For the
Halo~13 MW+LMC, the smallest mass splitting $\delta_m=475~\mathrm{keV}$ gives ${N_{\rm SB}^{\rm raw}}/{N_{\rm ROI}}\simeq 3.39$. The high-energy
sideband therefore provides an important additional test of this scenario,
but does not by itself establish an exclusion without the corresponding
detector acceptance.

{\it Acknowledgments} We thank Yongheng Xu for helpful discussions. L.W. is supported by the National Natural Science Foundation of China (NSFC) under grant No. 12335005 and No.~12275134. Y.Z. is supported by the Natural Science Foundation of Henan Province (No. 262300421233).

\bibliography{apssamp}

@article{Bertone:2004pz,
    author = "Bertone, Gianfranco and Hooper, Dan and Silk, Joseph",
    title = "{Particle dark matter: Evidence, candidates and constraints}",
    eprint = "hep-ph/0404175",
    archivePrefix = "arXiv",
    reportNumber = "FERMILAB-PUB-04-047-A",
    doi = "10.1016/j.physrep.2004.08.031",
    journal = "Phys. Rept.",
    volume = "405",
    pages = "279--390",
    year = "2005"
}

@article{Beneke:2022eci,
    author = "Beneke, Martin and Urban, Kai and Vollmann, Martin",
    title = "{Matching resummed endpoint and continuum {\ensuremath{\gamma}}-ray spectra from dark-matter annihilation}",
    eprint = "2203.01692",
    archivePrefix = "arXiv",
    primaryClass = "hep-ph",
    reportNumber = "TUM-HEP-1390/22",
    doi = "10.1016/j.physletb.2022.137248",
    journal = "Phys. Lett. B",
    volume = "834",
    pages = "137248",
    year = "2022"
}

@article{HESS:2026ila,
    author = "Aharonian, F. and others",
    collaboration = "H.E.S.S.",
    title = "{Search for Gamma-Ray Spectral Lines from Dark Matter Annihilation with the H.E.S.S. Inner Galaxy Survey}",
    eprint = "2608.07234",
    archivePrefix = "arXiv",
    primaryClass = "astro-ph.HE",
    doi = "10.1103/d8tj-55kc",
    journal = "Phys. Rev. Lett.",
    volume = "137",
    number = "9",
    pages = "091002",
    year = "2026"
}

@article{Urban:2021cdu,
    author = "Urban, Kai",
    title = "{NLO electroweak potentials for minimal dark matter and beyond}",
    eprint = "2108.07285",
    archivePrefix = "arXiv",
    primaryClass = "hep-ph",
    reportNumber = "TUM-HEP-1361/21",
    doi = "10.1007/JHEP10(2021)136",
    journal = "JHEP",
    volume = "10",
    pages = "136",
    year = "2021"
}

@article{Planck:2018vyg,
    author = "Aghanim, N. and others",
    collaboration = "Planck",
    title = "{Planck 2018 results. VI. Cosmological parameters}",
    eprint = "1807.06209",
    archivePrefix = "arXiv",
    primaryClass = "astro-ph.CO",
    doi = "10.1051/0004-6361/201833910",
    journal = "Astron. Astrophys.",
    volume = "641",
    pages = "A6",
    year = "2020",
    note = "[Erratum: Astron.Astrophys. 652, C4 (2021)]"
}

@article{Nagata:2014wma,
    author = "Nagata, Natsumi and Shirai, Satoshi",
    title = "{Higgsino Dark Matter in High-Scale Supersymmetry}",
    eprint = "1410.4549",
    archivePrefix = "arXiv",
    primaryClass = "hep-ph",
    reportNumber = "DESY-14-180, FTPI-MINN-14-37, IPMU14-0320",
    doi = "10.1007/JHEP01(2015)029",
    journal = "JHEP",
    volume = "01",
    pages = "029",
    year = "2015"
}

@article{Jungman:1995df,
    author = "Jungman, Gerard and Kamionkowski, Marc and Griest, Kim",
    title = "{Supersymmetric dark matter}",
    eprint = "hep-ph/9506380",
    archivePrefix = "arXiv",
    reportNumber = "SU-4240-605, UCSD-PTH-95-02, IASSNS-HEP-95-14, CU-TP-677",
    doi = "10.1016/0370-1573(95)00058-5",
    journal = "Phys. Rept.",
    volume = "267",
    pages = "195--373",
    year = "1996"
}

@article{Tucker-Smith:2001myb,
    author = "Tucker-Smith, David and Weiner, Neal",
    title = "{Inelastic dark matter}",
    eprint = "hep-ph/0101138",
    archivePrefix = "arXiv",
    reportNumber = "UCB-PTH-00-43, LBNL-47234, UW-PT-00-17",
    doi = "10.1103/PhysRevD.64.043502",
    journal = "Phys. Rev. D",
    volume = "64",
    pages = "043502",
    year = "2001"
}

@article{Bramante:2016rdh,
    author = "Bramante, Joseph and Fox, Patrick J. and Kribs, Graham D. and Martin, Adam",
    title = "{Inelastic frontier: Discovering dark matter at high recoil energy}",
    eprint = "1608.02662",
    archivePrefix = "arXiv",
    primaryClass = "hep-ph",
    reportNumber = "FERMILAB-PUB-16-301-T",
    doi = "10.1103/PhysRevD.94.115026",
    journal = "Phys. Rev. D",
    volume = "94",
    number = "11",
    pages = "115026",
    year = "2016"
}

@article{Nussinov:2009ft,
    author = "Nussinov, Shmuel and Wang, Lian-Tao and Yavin, Itay",
    title = "{Capture of Inelastic Dark Matter in the Sun}",
    eprint = "0905.1333",
    archivePrefix = "arXiv",
    primaryClass = "hep-ph",
    doi = "10.1088/1475-7516/2009/08/037",
    journal = "JCAP",
    volume = "08",
    pages = "037",
    year = "2009"
}

@article{Dessert:2022evk,
    author = "Dessert, Christopher and Foster, Joshua W. and Park, Yujin and Safdi, Benjamin R. and Xu, Weishuang Linda",
    title = "{Higgsino Dark Matter Confronts 14~Years of Fermi {\ensuremath{\gamma}}-Ray Data}",
    eprint = "2207.10090",
    archivePrefix = "arXiv",
    primaryClass = "hep-ph",
    reportNumber = "MIT-CTP/5454",
    doi = "10.1103/PhysRevLett.130.201001",
    journal = "Phys. Rev. Lett.",
    volume = "130",
    number = "20",
    pages = "201001",
    year = "2023"
}

@article{Catena:2018vzc,
    author = {Catena, Riccardo and Hellstr{\"o}m, Fredrik},
    title = "{New constraints on inelastic dark matter from IceCube}",
    eprint = "1808.08082",
    archivePrefix = "arXiv",
    primaryClass = "astro-ph.CO",
    doi = "10.1088/1475-7516/2018/10/039",
    journal = "JCAP",
    volume = "10",
    pages = "039",
    year = "2018"
}

@article{Beneke:2019gtg,
    author = "Beneke, Martin and Hasner, Caspar and Urban, Kai and Vollmann, Martin",
    title = "{Precise yield of high-energy photons from Higgsino dark matter annihilation}",
    eprint = "1912.02034",
    archivePrefix = "arXiv",
    primaryClass = "hep-ph",
    reportNumber = "TUM-HEP-1240/19",
    doi = "10.1007/JHEP03(2020)030",
    journal = "JHEP",
    volume = "03",
    pages = "030",
    year = "2020"
}

@article{CTAO:2024wvb,
    author = "Abe, S. and others",
    collaboration = "CTAO",
    title = "{Dark matter line searches with the Cherenkov Telescope Array}",
    eprint = "2403.04857",
    archivePrefix = "arXiv",
    primaryClass = "hep-ph",
    doi = "10.1088/1475-7516/2024/07/047",
    journal = "JCAP",
    volume = "07",
    pages = "047",
    year = "2024"
}

@misc{IceCube:2025fcu,
    author = "Abbasi, R. and others",
    collaboration = "IceCube",
    title = "{Search for High-Energy Neutrinos From the Sun Using Ten Years of IceCube Data}",
    eprint = "2507.08457",
    archivePrefix = "arXiv",
    primaryClass = "hep-ex",
    month = "7",
    year = "2025"
}

@article{Martin:2024ytt,
    author = "Martin, Stephen P.",
    title = "{Curtain lowers on directly detectable higgsino dark matter}",
    eprint = "2412.08958",
    archivePrefix = "arXiv",
    primaryClass = "hep-ph",
    doi = "10.1103/PhysRevD.111.075004",
    journal = "Phys. Rev. D",
    volume = "111",
    number = "7",
    pages = "075004",
    year = "2025"
}

@dataset{cherenkov_telescope_array_observatory_2021_5499840,
  author       = {Cherenkov Telescope Array Observatory and
                  Cherenkov Telescope Array Consortium},
  title        = {CTAO Instrument Response Functions - prod5 version
                   v0.1
                  },
  month        = sep,
  year         = 2021,
  publisher    = {Zenodo},
  version      = {v0.1},
  doi          = {10.5281/zenodo.5499840},
  url          = {https://doi.org/10.5281/zenodo.5499840},
}

@article{Rodd:2024qsi,
    author = "Rodd, Nicholas L. and Safdi, Benjamin R. and Xu, Weishuang Linda",
    title = "{CTA and SWGO can discover Higgsino dark matter annihilation}",
    eprint = "2405.13104",
    archivePrefix = "arXiv",
    primaryClass = "hep-ph",
    doi = "10.1103/PhysRevD.110.043003",
    journal = "Phys. Rev. D",
    volume = "110",
    number = "4",
    pages = "043003",
    year = "2024"
}

@article{Hisano:2006nn,
    author = "Hisano, Junji and Matsumoto, Shigeki and Nagai, Minoru and Saito, Osamu and Senami, Masato",
    title = "{Non-perturbative effect on thermal relic abundance of dark matter}",
    eprint = "hep-ph/0610249",
    archivePrefix = "arXiv",
    reportNumber = "KEK-TH-1111",
    doi = "10.1016/j.physletb.2007.01.012",
    journal = "Phys. Lett. B",
    volume = "646",
    pages = "34--38",
    year = "2007"
}

@article{Hisano:2004ds,
    author = "Hisano, Junji and Matsumoto, Shigeki. and Nojiri, Mihoko M. and Saito, Osamu",
    title = "{Non-perturbative effect on dark matter annihilation and gamma ray signature from galactic center}",
    eprint = "hep-ph/0412403",
    archivePrefix = "arXiv",
    reportNumber = "ICRR-REPORT-513-2004-11, YITP-04-73",
    doi = "10.1103/PhysRevD.71.063528",
    journal = "Phys. Rev. D",
    volume = "71",
    pages = "063528",
    year = "2005"
}

@article{Arkani-Hamed:2006wnf,
    author = "Arkani-Hamed, N. and Delgado, A. and Giudice, G. F.",
    title = "{The Well-tempered neutralino}",
    eprint = "hep-ph/0601041",
    archivePrefix = "arXiv",
    reportNumber = "CERN-PH-TH-2005-260",
    doi = "10.1016/j.nuclphysb.2006.02.010",
    journal = "Nucl. Phys. B",
    volume = "741",
    pages = "108--130",
    year = "2006"
}

@article{Blennow:2018xwu,
    author = "Blennow, Mattias and Clementz, Stefan and Herrero-Garcia, Juan",
    title = "{The distribution of inelastic dark matter in the Sun}",
    eprint = "1802.06880",
    archivePrefix = "arXiv",
    primaryClass = "hep-ph",
    reportNumber = "IFT-UAM-CSIC-18-019, ADP-18-4-T1052, IFT-UAM/CSIC-18-019, ADP-18-4/T1052",
    doi = "10.1140/epjc/s10052-018-5863-4",
    journal = "Eur. Phys. J. C",
    volume = "78",
    number = "5",
    pages = "386",
    year = "2018",
    note = "[Erratum: Eur.Phys.J.C 79, 407 (2019)]"
}

@article{Hisano:2004pv,
    author = "Hisano, Junji and Matsumoto, Shigeki and Nojiri, Mihoko M. and Saito, Osamu",
    title = "{Direct detection of the Wino and Higgsino-like neutralino dark matters at one-loop level}",
    eprint = "hep-ph/0407168",
    archivePrefix = "arXiv",
    reportNumber = "ICRR-REPORT-506-2004-4, YITP-04-39",
    doi = "10.1103/PhysRevD.71.015007",
    journal = "Phys. Rev. D",
    volume = "71",
    pages = "015007",
    year = "2005"
}

@dataset{bringmann_2024_11422081,
  author       = {Bringmann, Torsten and
                  Sæther Hatlen, Eirik and
                  Zaharijas, Gabrijela},
  title        = {Likelihoods for the CTA sensitivity to a dark
                   matter line signal from the Galactic centre (S.
                   Abe et al., 2024])
                  },
  month        = jun,
  year         = 2024,
  publisher    = {Zenodo},
  doi          = {10.5281/zenodo.11422081},
  url          = {https://doi.org/10.5281/zenodo.11422081},
}

@article{Jeong:2021bpl,
    author = "Jeong, Injun and Kang, Sunghyun and Scopel, Stefano and Tomar, Gaurav",
    title = "{WimPyDD: An object{\textendash}oriented Python code for the calculation of WIMP direct detection signals}",
    eprint = "2106.06207",
    archivePrefix = "arXiv",
    primaryClass = "hep-ph",
    reportNumber = "CQUeST-2021-0663, TUM-HEP 1343/21",
    doi = "10.1016/j.cpc.2022.108342",
    journal = "Comput. Phys. Commun.",
    volume = "276",
    pages = "108342",
    year = "2022"
}

@article{Helm:1956zz,
    author = "Helm, Richard H.",
    title = "{Inelastic and Elastic Scattering of 187-Mev Electrons from Selected Even-Even Nuclei}",
    doi = "10.1103/PhysRev.104.1466",
    journal = "Phys. Rev.",
    volume = "104",
    pages = "1466--1475",
    year = "1956"
}

@article{Lewin:1995rx,
    author = "Lewin, J. D. and Smith, P. F.",
    title = "{Review of mathematics, numerical factors, and corrections for dark matter experiments based on elastic nuclear recoil}",
    reportNumber = "RAL-TR-95-024",
    doi = "10.1016/S0927-6505(96)00047-3",
    journal = "Astropart. Phys.",
    volume = "6",
    pages = "87--112",
    year = "1996"
}

@article{Belanger:2001fz,
    author = "Belanger, G. and Boudjema, F. and Pukhov, A. and Semenov, A.",
    title = "{MicrOMEGAs: A Program for calculating the relic density in the MSSM}",
    eprint = "hep-ph/0112278",
    archivePrefix = "arXiv",
    reportNumber = "LAPTH-881-01",
    doi = "10.1016/S0010-4655(02)00596-9",
    journal = "Comput. Phys. Commun.",
    volume = "149",
    pages = "103--120",
    year = "2002"
}

@article{Slatyer:2015jla,
    author = "Slatyer, Tracy R.",
    title = "{Indirect dark matter signatures in the cosmic dark ages. I. Generalizing the bound on s-wave dark matter annihilation from Planck results}",
    eprint = "1506.03811",
    archivePrefix = "arXiv",
    primaryClass = "hep-ph",
    reportNumber = "MIT-CTP-4682",
    doi = "10.1103/PhysRevD.93.023527",
    journal = "Phys. Rev. D",
    volume = "93",
    number = "2",
    pages = "023527",
    year = "2016"
}

@misc{LZ:2026axp,
    author = "Akerib, D. S. and others",
    collaboration = "LZ",
    title = "{Search for dark matter particle interactions in an extended nuclear recoil energy window with the LUX-ZEPLIN (LZ) experiment}",
    eprint = "2609.02823",
    archivePrefix = "arXiv",
    primaryClass = "hep-ex",
    month = "9",
    year = "2026"
}

@article{ATLAS:2026hnb,
    author = "Aad, Georges and others",
    collaboration = "ATLAS",
    title = "{Search for long-lived charginos and {\ensuremath{\tau}}-sleptons using final states with a disappearing track in pp collisions at $ \sqrt{s}=13 $ TeV with the ATLAS detector}",
    eprint = "2603.08315",
    archivePrefix = "arXiv",
    primaryClass = "hep-ex",
    reportNumber = "CERN-EP-2026-044",
    doi = "10.1007/JHEP07(2026)152",
    journal = "JHEP",
    volume = "07",
    pages = "152",
    year = "2026"
}

@article{Fermi-LAT:2025gei,
    author = "Abdollahi, S. and others",
    collaboration = "Fermi-LAT, HAWC, H.E.S.S., MAGIC, VERITAS",
    title = "{Combined dark matter search towards dwarf spheroidal galaxies with Fermi-LAT, HAWC, H.E.S.S., MAGIC, and VERITAS}",
    eprint = "2508.20229",
    archivePrefix = "arXiv",
    primaryClass = "astro-ph.HE",
    doi = "10.1088/1475-7516/2026/03/035",
    journal = "JCAP",
    volume = "03",
    pages = "035",
    year = "2026"
}

@misc{Bisal:2026hpm,
    author = "Bisal, Subhadip and Chatterjee, Arindam and Das, Debottam and Pasha, Syed Adil and Puri, Rahul",
    title = "{Unveiling the Vanishing Higgsino-Nucleon Scattering in the MSSM at Next-to-Leading Order}",
    eprint = "2607.20588",
    archivePrefix = "arXiv",
    primaryClass = "hep-ph",
    month = "7",
    year = "2026"
}

@article{Catena:2015uha,
    author = "Catena, Riccardo and Schwabe, Bodo",
    title = "{Form factors for dark matter capture by the Sun in effective theories}",
    eprint = "1501.03729",
    archivePrefix = "arXiv",
    primaryClass = "hep-ph",
    doi = "10.1088/1475-7516/2015/04/042",
    journal = "JCAP",
    volume = "04",
    pages = "042",
    year = "2015"
}

@article{Asplund:2009fu,
    author = "Asplund, Martin and Grevesse, Nicolas and Sauval, A. Jacques and Scott, Pat",
    title = "{The chemical composition of the Sun}",
    eprint = "0909.0948",
    archivePrefix = "arXiv",
    primaryClass = "astro-ph.SR",
    doi = "10.1146/annurev.astro.46.060407.145222",
    journal = "Ann. Rev. Astron. Astrophys.",
    volume = "47",
    pages = "481--522",
    year = "2009"
}

@article{Alguero:2023zol,
    author = "Alguero, G. and Belanger, G. and Boudjema, F. and Chakraborti, S. and Goudelis, A. and Kraml, S. and Mjallal, A. and Pukhov, A.",
    title = "{micrOMEGAs 6.0: N-component dark matter}",
    eprint = "2312.14894",
    archivePrefix = "arXiv",
    primaryClass = "hep-ph",
    doi = "10.1016/j.cpc.2024.109133",
    journal = "Comput. Phys. Commun.",
    volume = "299",
    pages = "109133",
    year = "2024"
}

@article{Porod:2011nf,
    author = "Porod, W. and Staub, F.",
    title = "{SPheno 3.1: Extensions including flavour, CP-phases and models beyond the MSSM}",
    eprint = "1104.1573",
    archivePrefix = "arXiv",
    primaryClass = "hep-ph",
    doi = "10.1016/j.cpc.2012.05.021",
    journal = "Comput. Phys. Commun.",
    volume = "183",
    pages = "2458--2469",
    year = "2012"
}

@article{Porod:2003um,
    author = "Porod, Werner",
    title = "{SPheno, a program for calculating supersymmetric spectra, SUSY particle decays and SUSY particle production at e+ e- colliders}",
    eprint = "hep-ph/0301101",
    archivePrefix = "arXiv",
    reportNumber = "ZU-TH-01-03",
    doi = "10.1016/S0010-4655(03)00222-4",
    journal = "Comput. Phys. Commun.",
    volume = "153",
    pages = "275--315",
    year = "2003"
}

@misc{Pospelov:2026ewn,
    author = "Pospelov, Maxim and Ramani, Harikrishnan",
    title = "{Strong Constraints on Higgsino Dark Matter from Solar Capture}",
    eprint = "2609.02775",
    archivePrefix = "arXiv",
    primaryClass = "hep-ph",
    month = "9",
    year = "2026"
}

@misc{herrera_2023_10174172,
  author       = {Herrera, Yago and
                  Serenelli, Aldo},
  title        = {Standard Solar Models B23 / SF-III},
  month        = nov,
  year         = 2023,
  publisher    = {Zenodo},
  version      = {v1.1},
  doi          = {10.5281/zenodo.10174172},
  url          = {https://doi.org/10.5281/zenodo.10174172},
}

@article{Hahn:2000kx,
    author = "Hahn, Thomas",
    title = "{Generating Feynman diagrams and amplitudes with FeynArts 3}",
    eprint = "hep-ph/0012260",
    archivePrefix = "arXiv",
    reportNumber = "KA-TP-23-2000",
    doi = "10.1016/S0010-4655(01)00290-9",
    journal = "Comput. Phys. Commun.",
    volume = "140",
    pages = "418--431",
    year = "2001"
}

@article{Hahn:1998yk,
    author = "Hahn, T. and Perez-Victoria, M.",
    title = "{Automatized one loop calculations in four-dimensions and D-dimensions}",
    eprint = "hep-ph/9807565",
    archivePrefix = "arXiv",
    reportNumber = "UG-FT-87-98, KA-TP-7-1998",
    doi = "10.1016/S0010-4655(98)00173-8",
    journal = "Comput. Phys. Commun.",
    volume = "118",
    pages = "153--165",
    year = "1999"
}

@article{Smith-Orlik:2023kyl,
    author = "Smith-Orlik, Adam and others",
    title = "{The impact of the Large Magellanic Cloud on dark matter direct detection signals}",
    eprint = "2302.04281",
    archivePrefix = "arXiv",
    primaryClass = "astro-ph.GA",
    doi = "10.1088/1475-7516/2023/10/070",
    journal = "JCAP",
    volume = "10",
    pages = "070",
    year = "2023"
}

@article{ Asplund:2021solar,
	author = {{Asplund, M.} and {Amarsi, A. M.} and {Grevesse, N.}},
	title = {The chemical make-up of the Sun: A 2020 vision},
	DOI= "10.1051/0004-6361/202140445",
	url= "https://doi.org/10.1051/0004-6361/202140445",
	journal = {Astronomy \& Astrophysics},
	year = 2021,
	volume = 653,
	pages = "A141",
}

@article{Menon:2009qj,
    author = "Menon, Arjun and Morris, Rob and Pierce, Aaron and Weiner, Neal",
    title = "{Capture and Indirect Detection of Inelastic Dark Matter}",
    eprint = "0905.1847",
    archivePrefix = "arXiv",
    primaryClass = "hep-ph",
    reportNumber = "MCTP-09-14",
    doi = "10.1103/PhysRevD.82.015011",
    journal = "Phys. Rev. D",
    volume = "82",
    pages = "015011",
    year = "2010"
}

@article{Capdevilla:2024bwt,
    author = "Capdevilla, Rodolfo and Meloni, Federico and Zurita, Jose",
    title = "{Discovering Electroweak Interacting Dark Matter at Muon Colliders Using Soft Tracks}",
    eprint = "2405.08858",
    archivePrefix = "arXiv",
    primaryClass = "hep-ph",
    reportNumber = "FERMILAB-PUB-23-0832-T, DESY-24-069",
    doi = "10.1103/PhysRevLett.134.181802",
    journal = "Phys. Rev. Lett.",
    volume = "134",
    number = "18",
    pages = "181802",
    year = "2025"
}

@article{Mahbubani:2017gjh,
    author = "Mahbubani, Rakhi and Schwaller, Pedro and Zurita, Jose",
    title = "{Closing the window for compressed Dark Sectors with disappearing charged tracks}",
    eprint = "1703.05327",
    archivePrefix = "arXiv",
    primaryClass = "hep-ph",
    reportNumber = "CERN-TH-2017-054, TTP17-005, MITP/17-011",
    doi = "10.1007/JHEP06(2017)119",
    journal = "JHEP",
    volume = "06",
    pages = "119",
    year = "2017",
    note = "[Erratum: JHEP 10, 061 (2017)]"
}

@misc{Graham:2026ivn,
    author = "Graham, Peter W. and Ramani, Harikrishnan and Wong, Samuel S. Y.",
    title = "{Heavy-element paleodetectors for Higgsino dark matter}",
    eprint = "2606.05299",
    archivePrefix = "arXiv",
    primaryClass = "hep-ph",
    month = "6",
    year = "2026"
}

@article{Bisal:2024ezn,
    author = "Bisal, Subhadip and Chatterjee, Arindam and Das, Debottam and Pasha, Syed Adil",
    title = "{Radiative corrections to the direct detection of the Higgsino- and wino-like neutralino dark matter: Spin-dependent interactions}",
    eprint = "2410.18205",
    archivePrefix = "arXiv",
    primaryClass = "hep-ph",
    doi = "10.1103/PhysRevD.111.083003",
    journal = "Phys. Rev. D",
    volume = "111",
    number = "8",
    pages = "083003",
    year = "2025"
}

@article{Graham:2024syw,
    author = "Graham, Peter W. and Ramani, Harikrishnan and Wong, Samuel S. Y.",
    title = "{Enhancing direct detection of Higgsino dark matter}",
    eprint = "2409.07768",
    archivePrefix = "arXiv",
    primaryClass = "hep-ph",
    doi = "10.1103/PhysRevD.111.055030",
    journal = "Phys. Rev. D",
    volume = "111",
    number = "5",
    pages = "055030",
    year = "2025"
}

@article{Freese:2012xd,
    author = "Freese, Katherine and Lisanti, Mariangela and Savage, Christopher",
    title = "{Colloquium: Annual modulation of dark matter}",
    eprint = "1209.3339",
    archivePrefix = "arXiv",
    primaryClass = "astro-ph.CO",
    doi = "10.1103/RevModPhys.85.1561",
    journal = "Rev. Mod. Phys.",
    volume = "85",
    pages = "1561--1581",
    year = "2013"
}

@article{Piffl:2013mla,
    author = "Piffl, Til and others",
    title = "{The RAVE survey: the Galactic escape speed and the mass of the Milky Way}",
    eprint = "1309.4293",
    archivePrefix = "arXiv",
    primaryClass = "astro-ph.GA",
    doi = "10.1051/0004-6361/201322531",
    journal = "Astron. Astrophys.",
    volume = "562",
    pages = "A91",
    year = "2014"
}

@article{Baxter:2021pqo,
    author = "Baxter, D. and others",
    title = "{Recommended conventions for reporting results from direct dark matter searches}",
    eprint = "2105.00599",
    archivePrefix = "arXiv",
    primaryClass = "hep-ex",
    doi = "10.1140/epjc/s10052-021-09655-y",
    journal = "Eur. Phys. J. C",
    volume = "81",
    number = "10",
    pages = "907",
    year = "2021"
}

@misc{Rodd:2026tyn,
    author = "Rodd, Nicholas L. and Safdi, Benjamin R. and Slatyer, Tracy R. and Xu, Weishuang Linda",
    title = "{Confronting the Higgsino Interpretation of the LZ Event with the High-Energy Sideband}",
    eprint = "2609.04175",
    archivePrefix = "arXiv",
    primaryClass = "hep-ph",
    month = "9",
    year = "2026"
}

@article{Widmark:2017yvd,
    author = "Widmark, Axel",
    title = "{Thermalization time scales for WIMP capture by the Sun in effective theories}",
    eprint = "1703.06878",
    archivePrefix = "arXiv",
    primaryClass = "hep-ph",
    doi = "10.1088/1475-7516/2017/05/046",
    journal = "JCAP",
    volume = "05",
    pages = "046",
    year = "2017"
}

@article{Gould:1987ir,
    author = "Gould, Andrew",
    title = "{Resonant Enhancements in WIMP Capture by the Earth}",
    reportNumber = "SLAC-PUB-4226",
    doi = "10.1086/165653",
    journal = "Astrophys. J.",
    volume = "321",
    pages = "571",
    year = "1987"
}

@article{Spergel:1984re,
    author = "Spergel, D. N. and Press, W. H.",
    title = "{Effect of hypothetical, weakly interacting, massive particles on energy transport in the solar interior}",
    doi = "10.1086/163336",
    journal = "Astrophys. J.",
    volume = "294",
    pages = "663--673",
    year = "1985"
}

@article{Press:1985ug,
    author = "Press, William H. and Spergel, David N.",
    editor = "Srednicki, M. A.",
    title = "{Capture by the sun of a galactic population of weakly interacting massive particles}",
    doi = "10.1086/163485",
    journal = "Astrophys. J.",
    volume = "296",
    pages = "679--684",
    year = "1985"
}

@misc{Smirnov:2026aqk,
    author = "Smirnov, Juri and Griffith, Spencer and Beacom, John F.",
    title = "{Inelastic Signatures of Electroweak Dark Matter}",
    eprint = "2609.04144",
    archivePrefix = "arXiv",
    primaryClass = "hep-ph",
    month = "9",
    year = "2026"
}

\clearpage
\onecolumngrid
\hypersetup{hidelinks}

\begin{center}
{\large\bfseries Supplemental Material:\\
higgsino Dark Matter from LZ Nuclear Recoil to
Fermi-LAT Gamma Rays\par}
\vspace{1.3em}
Lei Wu$^{1}$, Yang Zhang$^{2}$, and Bin Zhu$^{3}$\\[0.35em]
{\itshape
$^{1}$Department of Physics and Institute of Theoretical Physics,\\
Nanjing Normal University, Nanjing 210023, China\\
$^{2}$Center for Theoretical Physics, Henan Normal University, Xinxiang 453007, China\\
$^{3}$School of Physics, Yantai University, Yantai 264005, China\par}
\end{center}
\vspace{1.5em}

\renewcommand{\theequation}{S.\arabic{equation}}
\renewcommand{\thefigure}{S\arabic{figure}}
\renewcommand{\thetable}{S\arabic{table}}
\setcounter{equation}{0}
\setcounter{figure}{0}
\setcounter{table}{0}
\renewcommand{\theHequation}{supp.\arabic{equation}}
\renewcommand{\theHfigure}{supp.\arabic{figure}}
\renewcommand{\theHtable}{supp.\arabic{table}}

This Supplemental Material provides the model construction and
computational details underlying the likelihood analysis and complementary
tests. We first show how weak gaugino admixtures generate the sub-MeV higgsino
splitting in the MSSM and give a representative benchmark. We then describe
the LZ and Fermi-LAT likelihood reconstructions, with the relevant halo and
response inputs specified in the corresponding sections. The remaining
sections discuss the IceCube solar-neutrino search, the
H.E.S.S. and CTAO gamma-ray searches, and the treatment of dwarf-spheroidal,
CMB, and collider constraints.

\section{higgsino Dark Matter in the MSSM}
\label{sec:mssm-realization}

The sub-MeV neutral-state splitting required by the LZ recoil arises
from weak gaugino admixtures in a nearly pure MSSM higgsino.
We take real
electroweakino parameters and the gauge-eigenstate basis
\(\psi^0=(-i\widetilde B,-i\widetilde W^3,
\widetilde H_d^0,\widetilde H_u^0)\), for which the neutralino mass matrix is
\begin{equation}
 {\cal M}_N=
 \begin{pmatrix}
 M_1 & 0 & -m_Zs_Wc_\beta & m_Zs_Ws_\beta \\
 0 & M_2 & m_Zc_Wc_\beta & -m_Zc_Ws_\beta \\
 -m_Zs_Wc_\beta & m_Zc_Wc_\beta & 0 & -\mu \\
 m_Zs_Ws_\beta & -m_Zc_Ws_\beta & -\mu & 0
 \end{pmatrix}.
 \label{eq:neutralino-matrix-supp}
\end{equation}
Here \(s_W=\sin\theta_W\), \(c_W=\cos\theta_W\),
\(s_\beta=\sin\beta\), and \(c_\beta=\cos\beta\). The mass
eigenstates \(\widetilde\chi_i^0=N_{ij}\psi_j^0\) are defined by
\begin{equation}
 N^*{\cal M}_NN^\dagger
 ={\rm diag}\!\left(
 m_{\widetilde\chi_1^0},m_{\widetilde\chi_2^0},
 m_{\widetilde\chi_3^0},m_{\widetilde\chi_4^0}\right),
 \qquad m_{\widetilde\chi_i^0}>0 ,
 \label{eq:neutralino-diagonalization-supp}
\end{equation}
with the states ordered by increasing mass. For the lightest
pseudo-Dirac pair, we define the higgsino fraction and neutral-current
couplings by
\begin{equation}
 f_H=|N_{13}|^2+|N_{14}|^2,\qquad
 D_{ij}=N_{i3}N_{j3}^*-N_{i4}N_{j4}^* .
 \label{eq:higgsino-fraction-z-coupling-supp}
\end{equation}
The pure-higgsino limit gives \(f_H\to1\), \(D_{11}\to0\), and
\(|D_{12}|\to1\). It therefore suppresses the diagonal neutral current while
retaining the weak-strength transition
\(\widetilde\chi_1^0 A\to\widetilde\chi_2^0 A\)~\cite{Nagata:2014wma}.

Away from a gaugino--higgsino crossing, the leading electroweak-mixing
expansion separates the bino and wino contributions to the neutral
splitting~\cite{Nagata:2014wma,Martin:2024ytt},
\begin{equation}
 \begin{split}
 \delta_m^{\rm tree}
 &\equiv
 m_{\widetilde\chi_2^0}^{\rm tree}
 -m_{\widetilde\chi_1^0}^{\rm tree}
 \simeq m_Z^2|C_B+C_W|,\\
 C_B&=s_W^2
 \frac{M_1+\mu\sin2\beta}{M_1^2-\mu^2},\qquad
 C_W=c_W^2
 \frac{M_2+\mu\sin2\beta}{M_2^2-\mu^2},
 \end{split}
 \label{eq:higgsino-splitting-components-supp}
\end{equation}
valid for
\(|M_i^2-\mu^2|\gg m_Z\max(|M_i|,|\mu|)\). Opposite-sign \(C_B\) and
\(C_W\) can partially cancel at finite gaugino masses, whereas same-sign
contributions require both gauginos to decouple. For
\(|M_{1,2}|\gg|\mu|\), these two regimes reduce to
\begin{equation}
 M_1\simeq-\tan^2\theta_W\,M_2
 \quad\text{(cancellation)},\qquad
 \frac{s_W^2}{|M_1|}+\frac{c_W^2}{|M_2|}
 \simeq\frac{\delta_m^{\rm tree}}{m_Z^2}
 \quad\text{(common-sign decoupling)} .
 \label{eq:gaugino-limits-supp}
\end{equation}
A splitting of a few hundred keV therefore requires a characteristic
gaugino scale of order \(10^4\,{\rm TeV}\) in the common-sign decoupling regime,
but can occur at lower scales through opposite-sign cancellation.

The resulting small
neutral splitting, however, does not by itself ensure a neutralino LSP.
In the same heavy-gaugino limit,
\begin{equation}
 m_{\widetilde\chi_1^\pm}^{\rm tree}
 -m_{\widetilde\chi_1^0}^{\rm tree}
 \simeq \frac{m_Z^2}{2}\left[
 \left|\frac{s_W^2}{M_1}+\frac{c_W^2}{M_2}\right|
 +\operatorname{sgn}(\mu)\sin2\beta
 \left(\frac{s_W^2}{M_1}-\frac{c_W^2}{M_2}\right)\right].
 \label{eq:chargino-neutralino-ordering-supp}
\end{equation}
The cancellation suppresses the sum in the first term but not the difference
in the second, so the tree-level ordering can be inverted for smaller
\(\tan\beta\) and less-decoupled gauginos~\cite{Martin:2024ytt}. Viable
points must therefore satisfy \(m_{\widetilde\chi_1^\pm}^{\rm pole}>
m_{\widetilde\chi_1^0}^{\rm pole}\).

\begin{table}[tbp]
 \caption{\textbf{MSSM benchmark point.}}
 \label{tab:joint-benchmark-supp}
 \footnotesize
 \begin{ruledtabular}
 \begin{tabular}{@{}lclc@{}}
 Quantity & Value & Quantity & Value\\
 \hline
 \(\mu\) & \(-1.100~{\rm TeV}\) & \(\tan\beta\) & \(5.25\)\\
 \(M_1\) & \(-3.475~{\rm TeV}\) &
 \(M_2\) & \(+10.50~{\rm TeV}\)\\
 \(m_{\widetilde\chi_1^0}\) & \(1.10562~{\rm TeV}\) &
 \(\delta_m\) & \(370~{\rm keV}\)\\
 \(f_H\) & \(0.99995\) & \(|D_{12}|\) & \(0.99984\)\\
 \(m_{\widetilde\chi_1^\pm}-m_{\widetilde\chi_1^0}\) &
 \(164~{\rm MeV}\) & \(\Omega_\chi h^2\) & \(0.12362\)\\
 \(\langle\sigma v\rangle_0\) &
 \(8.0668\times10^{-27}~{\rm cm^3\,s^{-1}}\) &
 \({\rm BR}_{WW}\) & \(0.54442\)\\
 \(\tau_{\widetilde\chi_2^0}\) & \(2.62\times10^{-2}~{\rm s}\) &
 \({\rm BR}_{ZZ}\) & \(0.45447\)\\
 \(\sigma_{p,{\rm inel}}^0\) &
 \(6.7635\times10^{-41}~{\rm cm^2}\) &
 \(\sigma_{n,{\rm inel}}^0\) & \(7.6494\times10^{-39}~{\rm cm^2}\)\\
 \(\sigma_p^{\rm SI}\) & \(2.6985\times10^{-49}~{\rm cm^2}\) &
 \(\sigma_n^{\rm SI}\) & \(2.7593\times10^{-49}~{\rm cm^2}\)\\
 \(\sigma_p^{\rm SD}\) & \(1.0048\times10^{-45}~{\rm cm^2}\) &
 \(\sigma_n^{\rm SD}\) & \(7.7044\times10^{-46}~{\rm cm^2}\)\\
 \end{tabular}
 \end{ruledtabular}
\end{table}

The benchmark in Table~\ref{tab:joint-benchmark-supp} realizes the
finite-gaugino cancellation while remaining in the pure-higgsino limit. 
We generate it using
\textsc{SPheno}~4.0.5~\cite{Porod:2003um,Porod:2011nf} with input parameters
at \(Q=4.5~{\rm TeV}\). Its relic density, perturbative short-distance
annihilation inputs, and tree-level elastic and inelastic cross sections are
obtained with micrOMEGAs~7.1~\cite{Belanger:2001fz,Alguero:2023zol}.

\section{LZ likelihood reconstruction}
\label{subsec:lz-likelihood-supp}

The endothermic threshold associated with the splittings relevant to the LZ
candidate pushes the required incident speed close to the endpoint of the
laboratory distribution.
The recoil rate must therefore apply the velocity threshold of each xenon
isotope separately. The profiled-Maxwellian treatment uses the
truncated-Maxwellian SHM adopted by LZ, parameterized by the speed triplet
\((v_0,v_{\rm esc},v_E)\):
\begin{equation}
 \begin{split}
 f_{\rm gal}(\bm v)&=
 \frac{e^{-v^2/v_0^2}}{N_{\rm esc}\pi^{3/2}v_0^3}
 \Theta(v_{\rm esc}-v),\qquad
 N_{\rm esc}={\rm erf}(z)-\frac{2z}{\sqrt\pi}e^{-z^2},
 \quad z=\frac{v_{\rm esc}}{v_0},\\
 f_{\rm lab}(\bm v)&=f_{\rm gal}(\bm v+\bm v_E),\qquad
 \eta(v_{\min})=\int_{|\bm v|>v_{\min}}
 \frac{f_{\rm lab}(\bm v)}{|\bm v|}\,\mathrm{d}^3v .
 \end{split}
\label{eq:mean-inverse-speed-supp}
\end{equation}

The second halo treatment instead replaces the Maxwellian mean inverse speed
with the present-day, best-Sun, MW+LMC mean curve of Halo~13 from
Ref.~\cite{Smith-Orlik:2023kyl}. We use its tabulated
\(\eta_{13}(v_{\min})\) as a fixed template, interpolate between the tabulated
speeds, and set it to zero beyond the terminal speed.

For our MSSM parameter space, \(D_{12}\) is purely imaginary in the pseudo-Dirac limit. The leading transition current is consequently vectorlike, with strength
\(|{\rm Im}\,D_{12}|=|D_{12}|\). Integrating out the \(Z\) boson gives the
zero-momentum reference cross section
\begin{equation}
 \sigma_A^0=
 \frac{G_F^2\mu_{\chi A}^2}{2\pi}
 \left(Q_W^A\right)^2|D_{12}|^2,
 \qquad
 Q_W^A=(A-Z_A)-(1-4s_W^2)Z_A .
 \label{eq:xenon-transition-cross-section-supp}
\end{equation}
We evaluate the full-density pure-higgsino signal by setting
\(\rho_{\widetilde\chi_1^0}=\rho_{\rm LZ}\) and \(|D_{12}|=1\), where
\(\rho_{\rm LZ}=0.30~\mathrm{GeV\,cm^{-3}}\) is the local density adopted by
LZ. The recoil rate per unit detector mass is
\begin{equation}
 \frac{\mathrm{d}R}{\mathrm{d}E_R}=
 \frac{\rho_{\rm LZ}}{m_{\widetilde\chi_1^0}}
 \sum_A\xi_A\frac{\sigma_A^0}{2\mu_{\chi A}^2}
 F_{W,A}^2(E_R)\,
 \eta\!\left[v_{\min}(E_R)\right],
 \qquad F_{W,A}^2(0)=1 .
 \label{eq:xenon-recoil-rate-supp}
\end{equation}
Here \(\xi_A\) is the natural-xenon mass fraction. We use the normalized
one-body responses
\(F_{W,A}^2(q)=W_{M,A}(q)/W_{M,A}(0)\) from
\textsc{WimPyDD}~\cite{Jeong:2021bpl}, with Helm form factors only for the
rare \(^{124}{\rm Xe}\) and \(^{126}{\rm Xe}\) isotopes absent from the
tabulation~\cite{Helm:1956zz,Lewin:1995rx}. Since \(Q_W^A\) is already
included in \(\sigma_A^0\), no additional weak-charge factor is applied.

All detector-level inputs entering the following construction are taken from the LZ extended nuclear-recoil
analysis~\cite{LZ:2026axp}. Let
\(r(E_R)=\mathrm{d}R/\mathrm{d}E_R\), and denote the digitized final-ROI
efficiency by \(\epsilon_{\rm ROI}(E_R)\). The accepted signal yield and
the reconstructed-energy intensity are
\begin{align}
 N_s&={\cal E}_{\rm LZ}\int\mathrm{d}E_R\,
 \epsilon_{\rm ROI}(E_R)r(E_R),\nonumber\\
 I_s(E;b_E)&={\cal E}_{\rm LZ}\int\mathrm{d}E_R\,
 \epsilon_{\rm ROI}(E_R)r(E_R)
 G(E-E_R-b_E;\sigma_{\rm stat}),
 \label{eq:lz-signal-count-supp}
\end{align}
where \({\cal E}_{\rm LZ}\) is the reported exposure. The Gaussian contains
only the reported statistical uncertainty.
The calibration displacement \(b_E\) translates the spectrum coherently and
has the prior \(b_E=0\pm\sigma_{\rm scale}\).

The official LZ analysis employs a simultaneous unbinned extended profile
likelihood in the two-dimensional \(\{S1_c,\log_{10}S2_c\}\) space for the
science, prompt-veto, and delayed-veto samples. Since the complete component
distributions are not public, we construct a one-event extended likelihood in
reconstructed recoil energy for the candidate at
\(E_{\rm obs}=248~\mathrm{keV}\). We use
\({\cal E}_{\rm LZ}=2.84~\mathrm{tonne\,yr}\) and
\(\sigma_{\rm stat}=\sigma_{\rm scale}=23~\mathrm{keV}\). Up to factors
independent of the signal model, the likelihood is
\begin{equation}
 {\cal L}_{\rm 1evt}\propto
 e^{-N_s}\left[I_s(E_{\rm obs};b_E)+\beta\right],
 \label{eq:lz-one-event-likelihood-supp}
\end{equation}
where \(\beta\) is an effective background intensity at the candidate energy.
We determine it from the public \(\mathcal O_1^s\) local significance
\(Z_{\rm ref}=3.0\) at
\((m_{\rm ref},\delta_{\rm ref})=(1~\mathrm{TeV},300~\mathrm{keV})\) with
\(b_E=0\).
Writing \(f_{s,{\rm ref}}=I_{s,{\rm ref}}/N_{s,{\rm ref}}\) and
\(r_\beta=\beta/f_{s,{\rm ref}}\), the one-event profile gives
\begin{equation}
 Z_{\rm ref}^2=2\left[-1+r_\beta-\ln r_\beta\right].
\label{eq:lz-background-calibration-supp}
\end{equation}
This fixes \(r_\beta=0.00410\) and
\(\beta=3.57\times10^{-6}~\mathrm{keV}^{-1}\) for our reconstructed-energy
surrogate. The pure-higgsino normalization is kept fixed in the parameter fit.

For each halo treatment, we define the profiled statistic
\begin{equation}
 \widetilde q_{{\rm LZ}}(m_\chi,\delta_m)=
 \min_{\boldsymbol\theta,b_E}\left\{
 2\left[N_s-\ln\!\left(I_s(E_{\rm obs};b_E)+\beta\right)\right]
 +\pi(\boldsymbol\theta)
 +\left(\frac{b_E}{\sigma_{\rm scale}}\right)^2
 \right\},
\label{eq:lz-test-statistic-supp}
\end{equation}
where \(\boldsymbol\theta=\boldsymbol v=(v_0,v_{\rm esc},v_E)\) for the
profiled Maxwellian and is absent for the fixed Halo~13 template. The
Maxwellian penalty is
\begin{equation}
 \pi_{\rm Max}(\boldsymbol v)=
 \sum_{i=1}^{3}\left(\frac{v_i-v_{c,i}}{\sigma_{v_i}}\right)^2,
 \qquad
 \boldsymbol v_c=(238,544,254)~\mathrm{km\,s^{-1}},
 \quad
 \boldsymbol\sigma_v=(20,50,15)~\mathrm{km\,s^{-1}},
 \label{eq:lz-speed-penalty-supp}
\end{equation}
while \(\pi_{13}=0\). The \(20~{\rm km\,s^{-1}}\)
width covers the shift from the conventional \(v_0=220~{\rm km\,s^{-1}}\)
choice to updated values near \(238~{\rm km\,s^{-1}}\)~\cite{Baxter:2021pqo}.
The \(50~{\rm km\,s^{-1}}\) width reflects the spread in local escape-speed
determinations~\cite{Piffl:2013mla}. The \(15~{\rm km\,s^{-1}}\) width is the
characteristic annual variation of the laboratory speed induced by the
Earth's orbital motion~\cite{Freese:2012xd}. For each halo treatment, the
quantity entering Eq.~\eqref{eq:joint_likelihood} is
\begin{equation}
 q_{{\rm LZ}}=\widetilde q_{{\rm LZ}}
 -\min_{m_\chi,\delta_m}\widetilde q_{{\rm LZ}}.
 \label{eq:lz-relative-statistic-supp}
\end{equation}

As a closure check, profiling the energy scale gives local significances that
agree with the published \(\mathcal O_1^s\) values to within about
\(0.3\sigma\) over \(\delta_m=100\)--\(350~\mathrm{keV}\). The baseline uses
only the recoil-energy support of the public final-ROI efficiency. Assigning a
constant \(5\%\) or \(10\%\) acceptance to the unreported
\(350\)--\(590~\mathrm{keV}\) sideband shifts the preferred Halo~13 splitting
by only a few keV and is not included in the central surface.

\begin{figure}[tbp]
 \centering
 \includegraphics[width=0.49\textwidth]{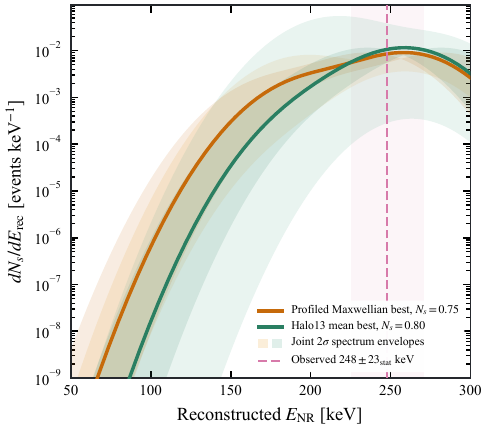}\hfill
 \includegraphics[width=0.49\textwidth]{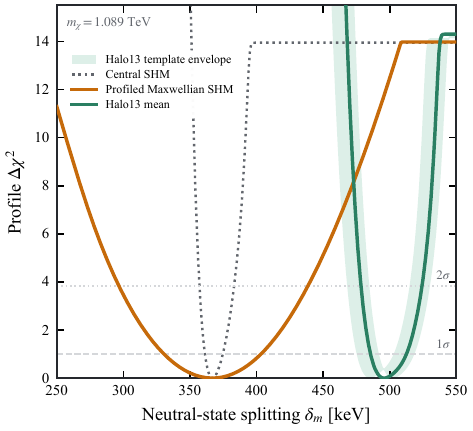}
 \caption{\textbf{LZ recoil reconstruction.}
 \textit{Left:} Accepted spectra at the profiled-Maxwellian and Halo~13 joint
 best fits, with envelopes over the joint \(2\sigma\) regions.
 \textit{Right:} One-dimensional LZ profiles at
 \(m_\chi=1.089\,\mathrm{TeV}\). The pale-green band spans the Halo~13
 lower, mean, and upper templates.}
 \label{fig:lz-spectrum-supp}
\end{figure}

Figure~\ref{fig:lz-spectrum-supp} summarizes the reconstructed recoil spectra
and the halo dependence of the LZ profile. In the left panel, the solid curves
show the accepted spectra at the two joint best-fit points. Their total yields
are \(N_s=0.75\) and \(0.80\), and both spectra peak within the statistical
uncertainty of the observed recoil. The broad bands are pointwise envelopes
over the corresponding joint two-dimensional
\(\Delta\chi^2\leq5.99\) regions, without unit-area normalization. The narrower
bands show coherent \(\pm1\sigma\) displacements of the reconstructed-energy
scale.

At fixed \(m_\chi=1.089~\mathrm{TeV}\), the central SHM and profiled-Maxwellian
curves in the right panel are both minimized near
\(\delta_m=367~\mathrm{keV}\). Profiling the velocity parameters substantially
broadens the Maxwellian profile. The Halo~13 mean template instead selects
\(\delta_m=496~\mathrm{keV}\), because its LMC-induced high-velocity tail keeps
the endothermic transition accessible at larger splittings. At sufficiently
large \(\delta_m\), the predicted signal vanishes and each profile approaches
its background-only plateau.

\section{Fermi-LAT likelihood reconstruction}
\label{sec:fermi-likelihood-supp}

The Fermi-LAT contribution is constructed from the public fourteen-year
Galactic-center likelihood for a higgsino continuum signal, provided in nine
angular annuli for an NFW density profile~\cite{Dessert:2022evk}.  For the
Galactic-center density, we instead adopt the Einasto profile
\begin{equation}
 \rho_{\rm Ein}(r)=\rho_{\odot,\gamma}
 \exp\!\left[-\frac{2}{\alpha}
 \left\{\left(\frac{r}{r_s}\right)^\alpha-
 \left(\frac{R_0}{r_s}\right)^\alpha\right\}\right],
 \label{eq:einasto-profile-supp}
\end{equation}
with \(\alpha=0.17\), \(r_s=20~{\rm kpc}\),
\(R_0=8.5~{\rm kpc}\), and
\(\rho_{\odot,\gamma}=0.40~{\rm GeV\,cm^{-3}}\).
For each annulus, we compute the \(J\) factor
\begin{equation}
 J_i^X=\int_{\Delta\Omega_i}\!\mathrm{d}\Omega
 \int_{\rm l.o.s.}\!\mathrm{d}s\,\rho_X^2[r(s,\Omega)],
 \qquad X\in\{\mathrm{Ein},\mathrm{NFW}\}.
 \label{eq:fermi-jfactor-supp}
\end{equation}
We map the expected Einasto normalization onto the signal-strength
coordinate of the public NFW likelihood through
\begin{equation}
 \ln {\cal L}^{\rm Ein}_F(m_\chi,\langle\sigma v\rangle)
 =\sum_{i=1}^{9}\ln {\cal L}^{\rm pub}_{F,i}\!\left(
 m_\chi,
 \langle\sigma v\rangle\frac{J_i^{\rm Ein}}{J_i^{\rm NFW}}
 \right).
 \label{eq:fermi-annulus-reweight-supp}
\end{equation}
The ratios are evaluated for the fixed Einasto benchmark above and the NFW
profile used in the public release.  Applying them annulus by annulus before
summing the likelihoods preserves the relative spatial normalization.

At each supplied mass, we sum the reweighted annular log likelihoods and
interpolate in the released signal-strength coordinate. We then interpolate
between the available higgsino mass templates without extrapolating beyond
their range.  The signal strength is fixed to the full-density
thermal-higgsino prediction \(\langle\sigma v\rangle_H\), including the
coupled electroweak annihilation channels and the Sommerfeld-enhanced
present-day rate~\cite{Hisano:2004ds,Hisano:2006nn,Dessert:2022evk}.  The
Fermi-LAT statistic \(q_F\) is then obtained from the signal-to-null likelihood
ratio defined in Table~\ref{tab:likelihood-contract}.
At \(m_\chi\simeq1.09~{\rm TeV}\), the continuum normalization used
in this ratio is
\(\langle\sigma v\rangle_H\simeq1.3\times10^{-26}~{\rm cm^3\,s^{-1}}\).
The public likelihoods already incorporate the diffuse-emission model,
point-source masks, and background-nuisance profiling. We modify only the
signal normalization assigned to each annulus through the corresponding
\(J\)-factor ratio.

\section{Solar capture, thermalization, and annihilation}
\label{sec:elastic-thermalization-supp}

The same off-diagonal \(Z\) transition that produces the LZ recoil also
captures halo higgsinos in the Sun through
\(\widetilde\chi_1^0 A\to\widetilde\chi_2^0 A\)~\cite{Nussinov:2009ft,Menon:2009qj,
Catena:2018vzc,Pospelov:2026ewn}.
Solar gravitational acceleration keeps this endothermic channel open on
heavy nuclei across most of the region of interest. The excited state then
decays rapidly to \(\widetilde\chi_1^0\). Residual elastic scattering determines whether
the captured population subsequently reaches a compact thermal distribution
in the solar core. We treat the thermal and nonthermal populations separately
below.

All three IceCube curves in Fig.~\ref{fig:icecube-exclusion} use the mass and
annihilation inputs in Table~\ref{tab:joint-benchmark-supp}, the Halo~13
MW+LMC mean speed distribution~\cite{Smith-Orlik:2023kyl}, and
\(\rho_{\widetilde\chi_1^0}=0.30~\mathrm{GeV\,cm^{-3}}\). We reconstruct the incident
speed distribution from the tabulated mean inverse speed according to
\begin{equation}
 g_{13}(u)=-u\frac{\mathrm d\eta_{13}(u)}{\mathrm du}.
 \label{eq:halo13-solar-speed-supp}
\end{equation}
We use \(g_{13}(u)\) directly as the solar-frame input and apply no additional
velocity boost.
The solar structure and abundances are taken from the B23/SF-III AGSS09
model~\cite{herrera_2023_10174172,Asplund:2009fu}. We use the
\(W_M\)-hybrid nuclear response~\cite{Catena:2015uha}, with Helm form
factors when a tabulated response is unavailable. The inelastic capture and
rescattering kernels use the vector-higgsino normalization in
Eq.~\eqref{eq:xenon-transition-cross-section-supp}~\cite{Pospelov:2026ewn}.

Our orbit-resolved Monte Carlo constructs the capture source by sampling the
incident halo speed \(u\), scattering radius \(r\), recoil energy \(E_R\), and
angular momentum \(L\). Gravitational focusing gives
\(w^2=u^2+v_{\rm esc}^2(r)\). We retain an inelastic event only when
\(w^2\geq2\delta_m/\mu_{\chi A}\) and the outgoing particle is gravitationally
bound. Its specific orbital energy is updated according to
\(E_f=E_i-E_R/m_\chi-\delta_m/m_\chi\), and each accepted event is assigned to
an \((E,L)\) orbit bin. The same differential cross section is used to
construct the subsequent inelastic transition kernel between bound orbits.
The capture integral uses \(1.2\times10^4\) samples for each nuclear species
and incident-state bin.

For the elastic benchmark, the solar evolution uses
\(\sigma_p^{\rm SI}=2.932\times10^{-47}~\mathrm{cm^2}\) and
\(\sigma_p^{\rm SD}=3.322\times10^{-45}~\mathrm{cm^2}\), including the
radiative contributions to the elastic amplitudes. At
\(\delta_m=370~\mathrm{keV}\), more than
\(99.9\%\) of the surviving captured particles reach the thermal reservoir.
Their median thermalization time is \(1.4\)--\(1.6~\mathrm{Myr}\), and the
equilibration time is about \(15.5~\mathrm{Myr}\), both well below the solar
age. We obtain the thermalization fraction and time distribution by following
successive inelastic and elastic collisions until each trajectory enters the
thermal reservoir or leaves the bound population. The thermal component has
\(V_{\rm eff}^{\rm th}=1.8779\times10^{26}~\mathrm{cm^3}\). Only after this
thermalization check do we use the equilibrium population equation
\begin{equation}
 \frac{\mathrm{d}N_\chi}{\mathrm{d}t}
 =C_\odot-C_A N_\chi^2,
 \qquad
 \Gamma_A=\frac{1}{2}C_A N_\chi^2,
 \label{eq:solar-population-supp}
\end{equation}
where \(C_\odot\) is the capture rate and
\(C_A=\langle\sigma v\rangle_{\rm eff}/V_{\rm eff}\), with the effective
annihilation rate including the Sommerfeld enhancement. At the solar age,
\begin{equation}
 \Gamma_A(t_\odot)=\frac{C_\odot}{2}
 \tanh^2\!\left(\frac{t_\odot}{\tau_A}\right),
 \qquad
 \tau_A=(C_\odot C_A)^{-1/2}.
 \label{eq:solar-equilibrium-supp}
\end{equation}

We compare the mixed \(WW+ZZ\) prediction with the 10-year IceCube
pure-\(WW\) rate limit~\cite{IceCube:2025fcu} through the response-matched
ratio
\begin{equation}
 R_{WW+ZZ}^{\rm resp}(\delta_m)=
 \frac{\Gamma_{WW}(\delta_m)+\kappa_{ZZ}\Gamma_{ZZ}(\delta_m)}
 {\Gamma_{A,WW}^{90}},
 \qquad \kappa_{ZZ}=1.157,
 \qquad
 \Gamma_{A,WW}^{90}=1.7210\times10^{19}~\mathrm{s}^{-1}.
 \label{eq:icecube-rate-ratio-supp}
\end{equation}
The relative response \(\kappa_{ZZ}\) follows from the public IceCube \(WW\)
spectrum and effective area with the solar-propagated \(ZZ/WW\) spectral
transfer. For the thermal benchmark, the Sommerfeld-convolved fractions
\(f_{WW}=0.6097\) and \(f_{ZZ}=0.3897\) give
\(\Gamma_X=f_X\Gamma_A\). Keeping the remaining particle inputs fixed,
\(R_{WW+ZZ}^{\rm resp}=1\) gives the approximate elastic-case limit
\(\delta_m\gtrsim553~\mathrm{keV}\).

\begin{figure}[t]
 \centering
 \includegraphics[width=0.95\textwidth]{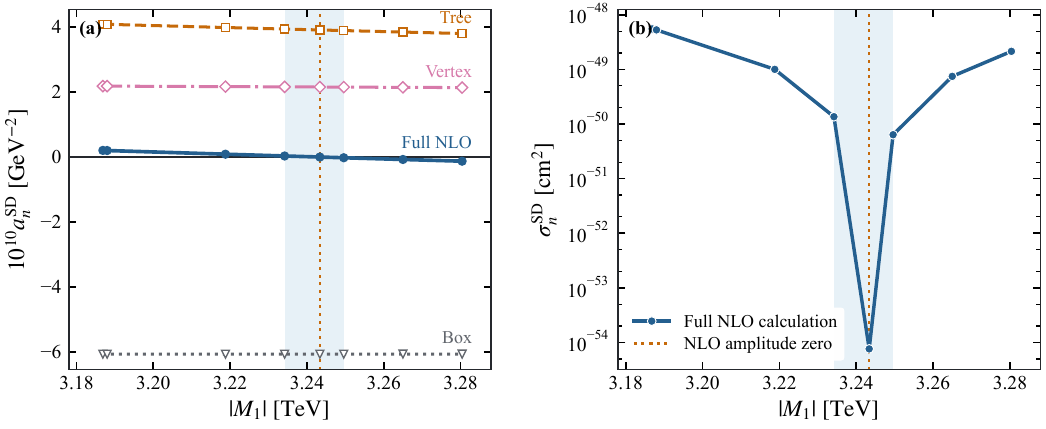}
 \caption{\textbf{Neutron spin-dependent cancellation.}
 Panel (a) shows the signed tree, electroweak-box, renormalized-vertex, and
 full amplitudes along a representative MSSM branch. Panel (b) shows the
 corresponding neutron SD cross section. }
 \label{fig:sd-cancellation-supp}
\end{figure}

The thermalized result depends on the residual elastic interactions. The
spin-independent nucleon cross section contains the coherent sum
\begin{equation}
 \sigma_N^{\rm SI}=\frac{4\mu_{\chi N}^2}{\pi}
 \left|\mathcal A_{N,h}^{\rm tree}
 +\mathcal A_{N,\rm EW}^{\rm loop}\right|^2 .
 \label{eq:coherent-si-supp}
\end{equation}
Tree--loop interference can suppress this rate, for example near the
top--stop threshold~\cite{Bisal:2026hpm}. For SD scattering, we calculate
the renormalized \(Z\)-vertex and electroweak-box amplitudes with
\textsc{FeynArts}~3.11~\cite{Hahn:2000kx} and
\textsc{FormCalc}~9.10~\cite{Hahn:1998yk} using the \textsc{MSSMCT} model,
with the loop integrals evaluated by
\textsc{LoopTools}~2.16. The resulting cancellation in the neutron amplitude
is shown in Fig.~\ref{fig:sd-cancellation-supp}.

To illustrate the cancellation, we follow a representative MSSM branch with
\(\mu=-1.10~\mathrm{TeV}\) and \(\tan\beta=4.22\). We vary \(M_1\) and retune
\(M_2\) at each point to maintain a sub-MeV neutral splitting, while keeping
the remaining inputs fixed. As \(|M_1|\)
increases, the gaugino
admixture and the asymmetry between the two neutral higgsino components
decrease, reducing the positive tree-level \(Z\)-exchange amplitude. The
negative electroweak-box amplitude is nearly unchanged, while the positive
vertex correction varies only mildly. At
\(M_1\simeq-3.24~\mathrm{TeV}\) and \(M_2\simeq6.81~\mathrm{TeV}\), the positive
tree and vertex terms balance the negative box contribution. The resulting amplitude is
\(a_n^{\rm SD}\simeq-2.4\times10^{-14}~\mathrm{GeV}^{-2}\), corresponding to
\(\sigma_n^{\rm SD}\simeq7.6\times10^{-55}~\mathrm{cm}^2\), which can be neglected.

In the limit of vanishing elastic scattering, solar nuclei still capture
halo higgsinos through \(\widetilde\chi_1^0 A\to\widetilde\chi_2^0 A\). Inelastic rescattering lowers
their orbital energies, but the particles do not settle into a compact
thermal distribution. They instead remain on extended solar orbits with
smaller spatial overlaps~\cite{Blennow:2018xwu}. We therefore do not impose
capture--annihilation equilibrium. For orbit states \(g\) with normalized
spatial profiles \(p_g(r)\), we define
\begin{equation}
 C_{A,gh}^{X}=\int\mathrm{d}V\,p_g(r)p_h(r)
 \left\langle\sigma v\right\rangle_{gh}^{X}(r),
 \qquad C_{A,gh}^{\rm tot}=\sum_X C_{A,gh}^{X},
 \label{eq:nonthermal-overlap-supp}
\end{equation}
where \(\langle\sigma v\rangle_{gh}^{X}(r)\) is the isotropic
relative-angle average of the channel-dependent Sommerfeld-enhanced
annihilation kernel for the orbit pair. For the benchmark, the populated
orbit pairs span \(S_{WW}=2.069\)--\(2.078\) and
\(S_{ZZ}=1.584\)--\(1.589\). Applied to
\(\langle\sigma v\rangle_0=8.0668\times10^{-27}~\mathrm{cm^3\,s^{-1}}\),
this gives a total Sommerfeld-enhanced cross section of approximately
\(1.5\times10^{-26}~\mathrm{cm^3\,s^{-1}}\).
We then evolve the orbit populations according to
\begin{align}
 \frac{\mathrm{d}N_g}{\mathrm{d}t}
 &=C_g+\sum_h\left(\Sigma_{hg}N_h-\Sigma_{gh}N_g\right)
 -\mathcal{E}_gN_g-N_g\sum_h C_{A,gh}^{\rm tot}N_h,\nonumber\\
 \Gamma_X
 &=\frac{1}{2}\sum_{g,h}C_{A,gh}^{X}N_gN_h .
 \label{eq:nonthermal-population-supp}
\end{align}
Here \(C_g\) continuously injects newly captured particles, \(\Sigma_{gh}\)
describes inelastic transitions between bound orbits, and \(\mathcal{E}_g\) is the
evaporation rate. Capture, orbital evolution, evaporation, and annihilation
are evolved simultaneously from the time of capture. The annihilation signal
is therefore determined by the time-dependent orbital overlaps and
relative-speed distributions rather than a final thermal reservoir. We
implement this evolution using 100 orbital-energy bins, 20 angular-momentum
bins, and 120 solar shells. The orbit overlaps and Sommerfeld convolution use
320 radial shells, and 128 scattering samples per initial orbit determine
the transition probabilities. We inject 5000 equal-weight macroparticles over
100 time intervals spanning the solar age and update the orbit-dependent
annihilation hazard from the population at the start of each interval. With
the central solar and nuclear inputs, \(R_{WW+ZZ}^{\rm resp}=1\) gives
\(\delta_m\gtrsim475~\mathrm{keV}\).

The limit is further sensitive to the solar abundances and nuclear response.
Our weakest treatment keeps the same Halo~13 distribution and density, sets
the elastic amplitudes to zero, and uses the same nonthermal evolution. We
adopt the individual \(2\sigma\) lower abundances of Fe, Zn, Ge, Pb, and U
from Ref.~\cite{Asplund:2021solar} within the B23/SF-III AGSS09 structure.
This gives \(\log\epsilon_{\rm Fe}=7.38\). Relative to the B23 abundances,
Fe is multiplied by \(0.75858\), while Zn, Ge, Pb, and U are multiplied by
\(0.79433\), \(0.58884\), \(0.56234\), and \(0.87096\), respectively. The
added heavy elements follow the normalized Ni radial profile. This setup gives
\(\delta_m\gtrsim466~\mathrm{keV}\).

\section{H.E.S.S. and CTAO line and endpoint searches}
\label{sec:hess-ctao-supp}

We place the H.E.S.S. and CTAO searches on a common higgsino-template
normalization \(\mu_\gamma\), with \(\mu_\gamma=1\) denoting the canonical
full-density thermal prediction.  For H.E.S.S., the observed monochromatic
line limit~\cite{HESS:2026ila} becomes
\begin{equation}
 \mu_{\gamma,95}^{\rm HESS}(m_\chi)=
 \frac{\langle\sigma v\rangle_{\rm line,95}^{\rm HESS}(m_\chi)}
 {\langle\sigma v\rangle_{\rm line}^{H}(m_\chi)},
 \label{eq:hess-normalization-supp}
\end{equation}
We compute the denominator with \textsc{DM$\gamma$Spec} using
the H.E.S.S.~line convention
\(\langle\sigma v\rangle_{\rm line}^{H}
=\langle\sigma v\rangle_{\gamma\gamma}^{H}
+\tfrac12\langle\sigma v\rangle_{\gamma Z}^{H}\), where the factor
\(1/2\) accounts for the single photon in the \(\gamma Z\) final state.  We
use the NLO Sommerfeld calculation at \(v=10^{-3}\), with a
\(355~{\rm MeV}\) charged splitting~\cite{Beneke:2019gtg,Urban:2021cdu,
Beneke:2022eci}.
The H.E.S.S. comparison therefore uses the published line-only limits for
the two choices of the solar Galactocentric distance, \(R_0=8.178\) and
\(8.5~{\rm kpc}\).

For CTAO, the same \textsc{DM$\gamma$Spec} input provides the complete
higgsino photon-number spectrum, including the resummed line, endpoint, and
continuum contributions.  For spatial ring \(p\) and reconstructed-energy bin
\(k\), the \(\mu_\gamma=1\) signal template is
\begin{equation}
 s_{pk}^{H}=
 \frac{J_pT_p}{8\pi m_\chi^2}
 \int\!\mathrm{d}E_t\,
 A_{\rm eff}(E_t,p)D_{pk}(E_t)
 \frac{\mathrm{d}\langle\sigma v\rangle_\gamma}{\mathrm{d}E_t},
 \label{eq:ctao-signal-supp}
\end{equation}
where \(J_p\) follows from the fixed Einasto profile in
Eq.~\eqref{eq:einasto-profile-supp}, \(A_{\rm eff}\) is the effective area,
\(D_{pk}\) is the energy-dispersion probability, and \(T_p\) is the exposure.
We use the public Prod5 South response
~\cite{cherenkov_telescope_array_observatory_2021_5499840}, four
\(0.5^\circ\) Galactic-center rings, the published \(3\times3\) pointing
pattern, and \(500~\mathrm{h}\) total exposure~\cite{CTAO:2024wvb}.
In each ring, we profile a local power-law background normalization and slope
together with a \(2.5\%\) correlated response nuisance, and determine the
one-sided 95\% limit from \(q=2.71\).

We determine the relative full-spectrum gain by applying this local
background-only Asimov construction to the higgsino
spectrum and a monochromatic line, and anchor its absolute normalization to
the released CTAO line profile through
\begin{equation}
 \mu_{\gamma,95}^{\rm Alpha,match}
 =\mu_{\gamma,95}^{\rm Alpha,loc}
 \frac{\langle\sigma v\rangle_{\gamma\gamma,95}^{\rm pub}}
 {\langle\sigma v\rangle_{\gamma\gamma,95}^{\rm loc}},
 \label{eq:ctao-response-match-supp}
\end{equation}
where the public and local line limits use the same two-photon normalization.
The ratio isolates the gain from the higgsino spectral shape, while the
public line limit fixes the absolute sensitivity
~\cite{CTAO:2024wvb,bringmann_2024_11422081}.

The released response and profile likelihood correspond to the Alpha
configuration.  The CTAO performance study quotes an approximate factor-of-two
improvement in line sensitivity for the full-scope Omega configuration under
the same search strategy~\cite{CTAO:2024wvb}.  We
therefore obtain the Omega curve from
\(\mu_{\gamma,95}^{\rm Omega,est}
=\mu_{\gamma,95}^{\rm Alpha,match}/2\).  The quoted improvement divides the
upper limit while leaving the predicted higgsino flux unchanged.

\section{Other experimental constraints}
\label{sec:other-constraints-supp}

For dwarf spheroidal galaxies, we compare the thermal-higgsino continuum with
the combined five-experiment $W^+W^-$ limits~\cite{Fermi-LAT:2025gei}.  The
$W^+W^-$ channel serves as a proxy for the mixed $W^+W^-$ and $ZZ$ higgsino
final state because their TeV-scale continuum spectra are similar.  We retain
the two published $J$-factor determinations of Geringer-Sameth et al. and
Bonnivard et al. At $m_\chi\simeq1.09~{\rm TeV}$, the predicted rate is $0.038$
and $0.168$ of the respective 95\% limits.

The CMB constraint is evaluated through
$p_{\rm ann}=f_{\rm eff}\langle\sigma v\rangle/m_\chi$, using
$f_{\rm eff}=0.20$ for the electroweak-boson higgsino spectrum and the same
Sommerfeld-enhanced annihilation rate used in the Fermi-LAT comparison
~\cite{Slatyer:2015jla}. At the same
reference mass, the prediction is $0.0073$ of the Planck bound
~\cite{Planck:2018vyg}.

The benchmark charged splitting, \(\delta_\pm=164~\mathrm{MeV}\), corresponds
to a chargino lifetime of approximately \(0.4~\mathrm{ns}\)
~\cite{Nagata:2014wma}. The lifetime-dependent ATLAS disappearing-track search
excludes pure-higgsino charginos up to \(720~\mathrm{GeV}\) near its maximum
sensitivity at \(\tau_{\widetilde\chi_1^\pm}\simeq1~\mathrm{ns}\)
~\cite{ATLAS:2026hnb}. The \(1.106~\mathrm{TeV}\) benchmark chargino therefore
lies above the exclusion reach.

\end{document}